\documentclass[12pt]{article}
\usepackage{color}

\usepackage{amsmath,amssymb}
\newcommand{\beq}{\begin{equation}}
\newcommand{\eeq}[1]{\label{#1}\end{equation}}
\newcommand{\bea}{\begin{eqnarray}}
\newcommand{\eea}[1]{\label{#1}\end{eqnarray}}
\renewcommand{\Im}{{\rm Im}\,}

\usepackage{setspace}
\def\simleq{\; \raise0.3ex\hbox{$< $\kern-0.75em
      \raise-1.1ex\hbox{$\sim$}}\; }
   \def\simgeq{\; \raise0.3ex\hbox{$> $\kern-0.75em
      \raise-1.1ex\hbox{$\sim$}}\; }
\usepackage{amsmath, amssymb,slashed,url}
\usepackage{amsthm}
\usepackage{graphicx,epsfig}
\usepackage{latexsym}
\def\tr{\mbox{tr}\,}
\def\Tr{\mbox{Tr}\,}
\begin{document}
\begin{titlepage}
\begin{center}

\vskip 4 cm

{\Large \bf  COHERENT QUBIT MEASUREMENT IN CAVITY-TRANSMON QUANTUM SYSTEMS  }

\vskip 1 cm

{Massimo Porrati$^{a}$ and Seth Putterman$^{b}$}

\vskip .75 cm

{$^a$ \em Center for Cosmology and Particle Physics, \\ Department of Physics, New York University, \\726 Broadway, New York, NY 10003, USA}

\vskip .75 cm

{$^b$ \em Department of Physics and Astronomy, \\ University of California Los Angeles, \\ Los Angeles, CA 90095-1547 USA}\end{center}

\vskip 1.25 cm

\begin{abstract}
\noindent  
A measurement of the time between quantum jumps implies the capability to measure the next
jump. During the time between jumps the quantum system is not evolving in a closed or unitary
manner. While the wave function maintains phase coherence it evolves according to a non-
hermitian effective hamiltonian. So under null measurement the timing of the next quantum jump
can change by very many orders of magnitude when compared to 
rates obtained by multiplying lifetimes with occupation probabilities obtained via unitary transformation.
 The theory developed in 1987 for
atomic fluorescence is here extended to transitions in transmon qubits. These systems differ from
atoms in that they are read out with a harmonic cavity whose resonance is determined by the
state of the qubit. We extend our analysis of atomic fluorescence to this infinite level system by
treating the cavity as a quantum system. We find that next photon statistics is highly non
exponential and when implemented will enable faster readout, such as on time scales shorter
than the decay time of the cavity. Commonly used heterodyne measurements are applied on time
scales longer than the cavity lifetime. The overlap between the next photon theory and the theory
of heterodyne measurement which are described according to the Stochastic Schr\"odinger Equation  is elucidated. 
In the limit of large dispersion the intrinsic error for next jump detection -at short time- tends to zero. Whereas for short time dyne detection the error remains finite for all values of dispersion.
\end{abstract}
\pagestyle{empty}
\end{titlepage}
\tableofcontents
\section{Introduction}\label{intro}
A quantum system that is simultaneously driven and observed will execute deterministic changes in the amplitudes of occupation of its various levels $|i \rangle$ that are interrupted by quantum jumps. Various types of quantum measurements are possible. These include observation of the emission from a transition between levels such as can be recorded on a photodetector in the case of fluorescence. When fluorescence measurements are very efficient one can then measure the time between emissions or so to speak the
\textit{next} quantum jump.  In order to measure the time $t_j$ between successive jumps one has to be able to measure that a jump has not happened for times $t < t_j$. For an atom driven to fluorescence by a laser a measurement of the next quantum jump implies the ability to determine that no fluorescent photon have been emitted during the interval between jumps. Consider two different interpretations of such a null measurement: a) as no photons are emitted the quantum system evolves in a unitary manner during this interval; there is at all times a probability for spontaneous decay determined by the lifetime of a level, and its occupation as determined by the unitary evolution;  b) a null measurement by the photodetector constitutes an interaction with the atom which changes its temporal evolution for $t<t_j$. 
We shall see that similarly to the {\em ab initio} calculations in~\cite{pp} a) is incorrect and that the experimentally measurable difference between these two models is enormous. The ability to make a null measurement leads to an irreversible yet phase coherent change in the evolution of the wave function for $t < t_j$. 
This ``weak'' measurement dramatically changes the temporal dynamics of the quantum system. In particular, a null measurement can lead to intervals of zero photon emission that exceed what would be calculated for the unitary evolution, model a), by many, many orders of magnitude~\cite{pp, erber-putterman}. We shall be especially interested in the discrepancy between a) and b) when the quantum system has more than 2 levels and there is a large variation in the lifetimes of these levels. 

As this insight is the building block for this work we first
review as part of the introduction~\ref{nextPhotonEmission} the case of fluorescence from a
driven 3 level atom. Our goal is to extend this perspective to multilevel quantum
systems, such as transmons, that are dispersively coupled to damped driven
resonators. Dispersive coupling occurs when the resonant frequency of a cavity
depends upon the quantum level which is occupied by the ``atom.'' In our approach
the transmon is a solid state device modelled as an atom with discrete levels. The
resonator is treated as a quantum -not classical- system. This approach is possible
because measurements are made at the photodetector. First consider the resonator
by itself. We find that for times short compared to the decay time $1/\kappa$ of the
cavity the probability that the next emitted photon will be detected in the interval
$[t,d+dt]$ is $[dW/dt] dt$  where $W$ is not Poisson distributed but instead is
$W\sim \exp(-\bar{n} \kappa^3 t^3/12)$
 where $\bar{n}$ is the steady state occupation number at resonant
drive. The simple recipe for understanding and calculating this case is presented in section~\ref{2.1}.

A purpose of this paper is to develop a bridge between single photon detection which is the characteristic measurement technique for ion traps~\cite{bhiw} and the heterodyne technique~\cite{wiseman, wiseman-milburn} which is characteristic of measurements of solid state qubits such as transmons. Both methods use photodetectors but in dyne detection the photodetector records a signal made by superimposing the photon escaping from the cavity to a coherent source. The dyne measurements are described by the Stochastic Schr\"odinger Equation [SSE]. 
   A recipe for interpreting and calculating the SSE is presented in section~\ref{comp} In
order to compare our next photon formalism with the SSE we evaluate its detailed
mathematical foundation in section~\ref{meas} following the analysis of Wiseman~\cite{wiseman}. Our
comparison of these two methods indicates that whereas the theory of single
photon detection is fundamental, the SSE as developed for heterodyne detection is
not fundamental. It is a coarse grained theory where the smallest time interval $\Delta t$
must be long enough to allow for the detection of many photons. So heterodyne
detection is not useful for time scales significantly shorter than $1/\kappa$. In comparison
we argue that detection of the next jump due to the recording of a single photon is
a fundamental process described by an effective non-Hermitian Hamiltonian. We
show, that if one post selects from the SSE those events where the heterodyne
current is equal to its maximum-likelihood value, then the system evolves exactly
as during a null measurement period for short times. An analysis of a protocol for
short time measurements is provided in section~\ref{2.3}
When the driven resonator is dispersively coupled to a transmon we find [section~\ref{2.4}] that there is a cavity induced lifetime that establishes a connection between
lifetimes of the transmon states and intrinsic cavity-transmon parameters. This
lifetime plays an essential role in determining the transition to coherently evolving
dark periods, as well as the length of the dark period in a transmon that is driven to
Rabi flopping. In section~\ref{lifetime} this new parameter is invoked regarding three levels
systems such as employed in experiments with multiple transmons~\cite{devoret}.
As mentioned, section~\ref{2} provides the higher level/practical approach to the
implementation and interpretation of the theory of the next quantum jump and its
relation to heterodyne detection. That we extracted a correct theory is justified by
its derivation from the established theory of continuous measurement which is
reviewed in section~\ref{meas}. Based upon the first principles theory, section~\ref{lind} includes
mathematical derivations of response times when the cavity is initially in the high-
field and low-field coherent state configurations. The high field occurs when the
frequency of the cavity drive matches the resonant frequency for the level occupied
by the transmon and the low field state occurs when the transmon is in its other
level so that the system is off resonance. We calculate the relaxation rates for
these cases as well as the lifetimes of the dark periods and their evolution during
the continuously observed dark state.
   
The analysis in Section~\ref{lind} allows for the description of photon dynamics when emission is measured relative to the coherent state as well as the ground state.  
Given the advantages of measurements based upon detecting the next single photon the issue arises as to whether this technique can be applied to solid state systems. For experiments on ion traps a photomultiplier tube is the instrument of choice because of the 
   $O(eV)$ energy of a single photon. For solid state qubits operating in the GHz domain the lower photon energy necessitates the use of heterodyne detection. However we note that attempts are being made to develop methods for detecting single radiofrequency photons~\cite{A,B,C,D}. With this opportunity in mind we have extended our next quantum jump formalism~\cite{pp} to solid state systems. We wish to emphasize that this is not a philosophical paper on quantum measurement. We deal with [sometimes non-intuitive] experimentally measurable consequences of quantum theory that have implications for qubit readout and when optimized could drive the design of new devices.

\subsection{Next photon emission from a 3 level atom}\label{nextPhotonEmission}
The key issues that we wish to generalize in this paper are contained in the behavior of an atom with 3 levels that is externally driven with a time dependent electric field. So we first review this case. Consider a 3 level atom which has 2 excited states $|B \rangle, |D \rangle$ with a common ground state $|G \rangle$; there is no transition between the excited states. The transitions $|G \rangle$ - $|B \rangle$; $|G \rangle$ - $|D \rangle$ are externally driven near resonance with oscillating electric fields. A photodetector measures with 100\% efficiency the photons emitted via spontaneous decay. For scenario a) the amplitudes $C_i(t)$ to be in $| i \rangle$ at time $t$ are given by the evolution of the externally driven system governed by a Hamiltonian $H$. The probability of photon emission from levels $i = B,D$ within an interval $dt$ at time $t$ is $\beta_i |C_i(t)|^2 dt$ where the lifetime of the levels: $\tau_i = 1/\beta_i$ are due to spontaneous decay. For an external driving field tuned closely to the transition energies: 
\begin{equation}
E = E_B \cos \left( \omega_{BG} t \right) + E_D \cos \left(\omega_{DG} + \Delta \right) t;
\end{equation} 
the rotating wave approximation [RWA] yields:
\begin{equation}
H_{BG}^* = H_{GB} = -\hbar \Omega_B^*; \quad H_{DG}^* = H_{GD} = -\hbar \Omega_D^* \exp(i \Delta t) 
\end{equation}
where: $\Omega_i$ are the Rabi flopping frequencies that are proportional to $E_i$ and, 
\begin{equation}
i \hbar \frac{dC_i}{dt} = H_{ij} C_j.
\end{equation} 
\par
The effects of null measurement become dramatic when the lifetimes are well separated such as when $\tau_B \sim 1\textrm{ ns}$; $\tau_D \sim 1\textrm{ s}$. So we consider such a case and note that in this limit one also has that $|\Omega_D/\Omega_B| \ll 1$ so that to leading order [taking $\Delta = |\Omega_B|$ for computational convenience]:
\begin{equation}
C_B(t) = \frac{1}{2} \left( \exp \left( i|\Omega_B|t \right) \cos \left( |\Omega_D| t / \sqrt{2} \right) - \exp \left( -i \Omega_B t \right) \right).
\end{equation}
Evolution of the 3 level atom according to this Hermitian Hamiltonian starts from the initial condition that an outgoing photon has been recorded at $t = 0$ at which time the atom resets to the ground state so that $C_B(0) = C_D(0) = 0$. In a time $t \sim 1/|\Omega_B|$ the probability to be in state $|B\rangle$ reaches an average value of $1/2$ and there is a strong probability that a photon with frequency $\omega_{BG}$ is emitted in about a ns. Following this emission the atom resets to $|G\rangle$ and the same process repeats.  The probability that after a reset no photons are emitted for a time $T$ or longer is: $W(t) = \exp \left( -\beta_B T/2 \right)$, where we consider the strong drive limit where $|\Omega_B| \gtrsim \beta_B$. In particular the probability of a dark period of length longer that $\tau_D$   is for the above physically achievable case: $W(1.\mathrm{s}) = \exp \left( - 5 \cdot 10^8 \right)$. As there are about $1/\tau_B$ attempts per second the chance of seeing a dark period start per second is $W(T)/\tau_B$ which is still incredibly small. 
\par
According to the principles of quantum theory the above analysis is way off when the measuring apparatus is capable of measuring the time between jumps. In fact the percentage of time the emission is dark $p_D$ can be order unity even in the limit where the illuminating fields are completely coherent [i.e. have zero bandwidth] and even in the limit where the strong transition is driven at saturation: $|\Omega_B| \gtrsim \beta_B$. Under the condition $|\Omega_D|/\beta_B = \epsilon \ll 1$ : 
\begin{equation}
\label{equation:percentage_of_dark}
p_D = \frac{F}{2+F}
\end{equation}
where $F= (1 + \beta_B \beta_D/ 4|\Omega_D|^2)^{-1}$. This enormous increase in probability comes about from the implementation of scenario b). When the intervals of no emission are measured, the amplitudes $C_i(t)$ must be reinterpreted as the amplitude for the atom to be in level '$i$' at time '$t$' subject to the observation that no photons have been emitted since the last recorded detection which we take to be $t=0$. The probability that no outgoing photons have been recorded in the interval $[0,t]$ is
\begin{equation}
W(t) = \sum |C_i(t)|^2.
\end{equation}
In view of spontaneous decay this quantity is no longer conserved. Some of the probability transfers into states with nonzero outgoing photons. So the effective Hamiltonian for the evolution of the $C_i(t)$ is no longer Hermitian. However, the Hamiltonian evolution of the null emission atom is a closed description. While the $C_i(t)$ are a source for states with outgoing photons there are no states that feed the $C_i(t)$. For the 3 level atom the closed non-Hermitian equations for the evolution of the wave function between quantum jumps takes the simple form:
\begin{equation}
\label{equation:equations_for_the_evolution}
\begin{gathered}
\frac{dC_D}{dt} = i\Omega_D C_G + \left( i \Delta_2 - \frac{\beta_D}{2} \right) C_D, \\
\frac{dC_B}{dt} = i\Omega_B C_G - \left( \frac{\beta_B}{2} \right) C_B, \\
\frac{dC_G}{dt} = i\Omega_B^* C_B + i \Omega_D^* C_D .
\end{gathered}
\end{equation}
In \eqref{equation:equations_for_the_evolution} the non-Hermitian terms proportional to $\beta_B, \beta_D$ account for the transfer of probability amplitude to states with nonzero numbers of outgoing photons. The amplitude for the atom to be in the strongly emitting level is:
\begin{equation}
\label{equation:amplitude_of_the_atom}
C_B(t) = i \sin \left( |\Omega_B|t \right) \exp \left( -\frac{\beta_B t}{4} \right) + 4 \frac{|\Omega_D|^2}{\beta_B^2} \exp \left( i |\Omega_B| t \right) \left( \exp \left( -\frac{\beta_B t}{4} \right) - \exp \left( -\beta_{\ell} t \right) \right),
\end{equation}
where the long time scale is determined by
\begin{equation}
\label{equation:long_time_scale}
\beta_{\ell} = \frac{1}{2} \beta_D + \frac{2 |\Omega_D|^2}{\beta_B} \ll \beta_B.
\end{equation}
The separation of time scales in \eqref{equation:amplitude_of_the_atom}, \eqref{equation:long_time_scale} presents the opportunity to observe the system on a time scale $T$ such that $1/\beta_B \ll T < 1/\beta_{\ell}$. If there are no jumps during this time $T$ the wave function is projected onto the slowly evolving term proportional to $\exp ( - \beta_{\ell} t)$. For $1/\beta_B \ll T < t$ this implies: 
\begin{equation}
\label{equation:wave_function_slow}
\begin{gathered}
C_G(t) = C_B(t) = 2 i \epsilon f \exp \left( i |\Omega_B| t \right) \exp \left( -\beta_{\ell} (t-T) \right), \\
C_D(t) = f \exp \left( i |\Omega_B| t \right) \exp \left( -\beta_{\ell} (t-T) \right),
\end{gathered}
\end{equation}
where: $f = 1/ \sqrt{1 + 8\epsilon^2}$ is a normalization. Once the wave function takes the form \eqref{equation:wave_function_slow} the time to the next jump is determined by $\tau_{\ell} = 1/ \beta_{\ell}$. For example if no jump is observed for a time $T \sim 12 \tau_B$ the probability of observing a dark period of length $\tau_{\ell}$ is order unity. For the physical example above the observation of no emission from the strong transition for a time of about 12ns leads to a dark period of about 1.sec. The probability of not seeing an emission, after a reset, for 12 lifetimes of the strong transition is $\exp(-6)$ which is small. However, it is huge compared to the case of unitary evolution a).  As there are many resets per second the probability of a dark period is now large and the percentage of time dark can be order unity. 
\par
The observation of a long dark period is \underline{not} due to the absorption of a photon from the light source that is tuned to the slow  $|G \rangle$ - $|D \rangle$ transition. During the dark period the atom is not shelved in $|D \rangle$ but is in an evolving superposition of all 3 levels. A large fraction of dark periods end with a reset to $|G \rangle$ due to a photon being emitted from level $|B \rangle$! In Equation \eqref{equation:percentage_of_dark} $\Gamma$ is the percentage of long dark periods that terminate with the emission of a photon from level $|B \rangle$. We see that an absorption event does not precede the emission from $|D \rangle$. For absorption to preceed emission requires $\Gamma = 0$ which is a contradiction to the existence of a dark period as \eqref{equation:percentage_of_dark} would then imply that $p_D = 0$.  
\par
During the time between jumps when there has been no emission for $ t \gg 1/\beta_B$ the wave function is known and is evolving coherently according to \eqref{equation:wave_function_slow}. During this interval there is a window of time $\sim \tau_{\ell}$ when the wave function can be modified by changing the electric field so as to return $C_G$ to unity and suppress the next quantum jump \cite{devoret}. 

\section{Photon measurement in a coupled transmon-resonator system}\label{2}
\subsection{Next photon emission from a damped driven harmonic resonator}\label{2.1}
A method to readout the state of a two level quantum system [or qubit] is to couple it to a harmonic cavity in such a way that the resonant frequency of the cavity depends upon which level is occupied \cite{blais}. The response of the resonator-cavity to a driving field will then be sensitive to the state of the qubit, and become a measuring device for the qubit state. Here we treat the resonator as a quantum system in parallel with our calculations for the 
3 level atom in the previous section. In contrast to AMO qubits the method for reading out a resonator involves coupling to an infinity of levels. This is the practical method for reading out the state of transmon qubits.
 We calculate how the time to the next quantum jump of the resonator determines the state of the resonator and then also the qubit. Treating the coupled resonator as a quantum as compared to a classical system may be relevant to technological improvements in qubit readout where it is acknowledged that 
 ``Reducing the time required to distinguish qubit states with high fidelity is a critical goal in quantum-information science,'' \cite{walter}.  Next jump detection enables a measurement of short time response and might then have advantages over heterodyne measurements which are generally averaged over time scales longer than the lifetime $1/\kappa$ of the cavity/resonator. 
\par
Damping is an essential aspect of the quantum dynamics of the next jump. In the previous section we saw that spontaneous decay affected the evolution of the wave function in such a way as to enable the observation of extended periods of darkness from a driven atom. Spontaneous decay can be calculated from first principles and so the equations \eqref{equation:equations_for_the_evolution} for the evolution of the atom between photodetection events can be derived by deductive mathematics \eqref{mm35} and~\cite{pp}. 
Equation~\eqref{equation:equations_for_the_evolution} is then supplemented with the probability
\beq
D(t) dt = -{dW(t)\over dt} dt
\eeq{njr}
that the next jump is recorded between $t$ and $t+dt$. Equations~(\ref{equation:equations_for_the_evolution},\ref{njr})
were derived in~\cite{pp}, they constitute an extension of the theory of Srinivas and Davies~\cite{srida} to non-stationary processes
such as occur in the presence of a continuous drive. Srinivas and Davies~\cite{srida} showed that in the presence of a dissipative 
environment the evolution of the wave function itself can be described by an effective Hamiltonian during an interval when no counts are recorded. Their analysis assumed stationarity and was applied under conditions of free decay which is a Poisson process. The long dark periods and separation of time scales observed in the 3 level system do not appear under conditions of free decay or stationarity. The theory of Srinivas and Davies (SD) for the evolution of the density matrix matches Lindblad~\cite{lind} who also imposed stationarity in the presence of a dissipative environment.\footnote{We do not understand why this condition was so prominently imposed. Our first principles analysis~\cite{pp} and Appendix B as well as recent work on the Lindblad 
equation~\cite{pearle,weinberg}  suggest that in the general case one simply includes the time dependent potentials in the effective Hamiltonian.}
For transmon qubits the resonator decay time is phenomenological and is not derived from first principles. The transmon dynamics is obtained by supplementing
the decay process with time dependent excitations as in~\cite{pp}. This is equivalent to applying the Lindblad/SD formalism for the evolution of the density matrix/wave function with time dependent 
terms included in the effective Hamiltonian~\cite{pp}. A time dependent effective Hamiltonian evolution interspersed by wave function 
collapse was used by Tian and Carmichael~\cite{tiancarm}  to study a single atom in a two level cavity. They call this extension of the 
Lindblad/SD theory ``quantum trajectories.'' Below we calculate the trajectory for a two level atom coupled to cavity modes.
 
The Hamiltonian for a lossless driven resonator that is dispersively coupled to a transmon with ground state $|G \rangle$ and excited state $|B \rangle$ is modeled following~\cite{devoret} as:
\begin{equation}
\label{equation:resonator_hamiltonian}
H_R = \hbar \chi \left( |B \rangle \langle B| - 1 \right) c^\dagger c - i \hbar \Gamma \left( e^{i \omega_D t}c - e^{-i \omega_D t} c^\dagger  \right) + \hbar \omega_R c^\dagger c,
\end{equation}
where the strength and frequency of the external driving field are $\Gamma$, $\omega_D$. When the atom is in $|G \rangle$ the resonant frequency of the cavity is $\omega_C = \omega_R - \chi$. We take $\chi < 0$ as in \cite{devoret} so that the cavity resonance is lower when $|B\rangle$ is occupied. The evolution of the density matrix $\rho$ of an observed quantum system is determined by the Lindblad operator~\cite{lind} $\mathcal{L}$ as extended to apply to non-stationary processes:
\begin{equation}
i \hbar \frac{d \rho}{d t} = \mathcal{L} \rho = H_{eff} \rho - \rho H_{eff}^\dagger + \hbar \mathcal{I} \rho,
\end{equation}
where $\mathcal{I} \rho$ describes the observation. When $\mathcal{I}\rho = i \kappa c \rho c^\dagger$; $\kappa \mathrm{tr} c^\dagger c \rho(t) dt$ is the probability to detect a quantum of the cavity mode $c$ in the time interval $[t, t+dt]$. As the resonator must be in some state its trace is a constant of the motion: $\mathrm{Tr} d\rho/dt = 0$. In order to meet this condition the effective Hamiltonian must depend on the damping $\kappa$ and become non-Hermitian with:
\begin{equation}
\label{equation:effective_resonator_hamiltonian}
H_{eff} = H_R - i \hbar \left( \frac{\kappa}{2} \right) c^\dagger c .
\end{equation}
An underlying assumption in the construction of $H_{eff}$ is that damping is due to the escape of photons which are then observed. In the case of fluorescence in AMO systems these are photons from spontaneous decay whereas for the transmons these might be photons that leave the resonant cavity through a `hole'  in the wall.

Consider now the initial condition where the “atom” /transmon is in the ground state and the drive is tuned to resonance so that $\omega_D = \omega_C$, then the evolution of the resonator up until the first quantum jump is given by:
\begin{equation}
\label{equation:ode2_for_cgn}
\frac{dC_{G,n}}{dt} = \Gamma \left( \sqrt{n} C_{G, n-1} - \sqrt{n+1} C_{G, n+1} \right) - \left( \frac{\kappa n}{2} \right) C_{G,n},
\end{equation}
where $C_{G,n}(t)$ [with $C_{G,n}(0) = \delta_{n,0}$] is the amplitude for the atom to be in G and the resonator to be in state $n$ at time '$t$' subject to the measurement that no outgoing photons have been emitted during the interval $[0,t]$. This is a situation where the resonator drive is turned on at $t=0$ to determine the state of the qubit. Here we are calculating the response given that the qubit is in G. A solution exists in the form:
\begin{equation}
\label{equation:solution_form}
C_{G,n}(t) = \frac{\exp \left( \beta(t) \right) \alpha(t)^n}{\sqrt{n!}},
\end{equation}
where $\alpha(t)$ is a free parameter [not the index of the ancilla Hilbert space appearing in
eqs.~(\ref{qm1}-\ref{qm4})!]. For these initial conditions:
\begin{equation}
\label{equation:solution_beta}
\beta(t) = -\frac{\kappa}{2} \bar{n} \left( t + \frac{2}{\kappa} \left(e^{- \kappa t/2} -1\right) \right),
\end{equation}
\begin{equation}
\label{equation:ode_beta}
\frac{d\beta}{dt} = -\Gamma \alpha,
\end{equation}
And generally:
\begin{equation}
\label{equation:infinite_sum}
W(t) = \sum \limits_{n=0}^{\infty} \left| C_{G,n} \right|^2 = \exp \left( \beta + \beta^* + |\alpha|^2 \right).
\end{equation}
 We have introduced the steady state occupation 
of the resonator, $\bar{n}$, so that $\Gamma = 
\kappa \sqrt{\bar{n}}/2$. The probability that the next jump has not occurred in time '$t$' is $W(t)$. The probability that the next jump 
will occur between $t$ and $t+dt$ is:
\begin{equation}
\label{equation:probability_of_not_next_jump}
D(t) dt = - \left( \frac{dW}{dt} \right) dt = \kappa \alpha \alpha^* W.
\end{equation}
Eqs.~(\ref{equation:ode2_for_cgn},\ref{equation:probability_of_not_next_jump}) 
for the evolution of the driven cavity take the place 
of~~(\ref{equation:equations_for_the_evolution},\ref{njr}) for multi-level fluorescence. 
For short times this case gives:
\begin{equation}
\label{equation:probability_of_not_next_jump_solution}
W \sim \exp \left( -\bar{n} \kappa^3 t^3 /12  \right)
\end{equation}
and the expected time to the next jump when $\bar{n} \gg 1$ is $\bar{t}_j \sim (3/\kappa \Gamma^2)^{1/3}$. For times long compared to the cavity lifetime [$\kappa t > 1$] the non-exponential behavior \eqref{equation:probability_of_not_next_jump_solution} switches to exponential response so that $W \to W_0 \exp (- \kappa \bar{n} t)$. For a strong drive, $\Gamma$, the time to the first jump can be less than the resonator lifetime. For this reason a measurement of the next jump can offer advantages over heterodyne measurements of the state of the qubit. 
\par
Consider next the case where the atom is reinitialized in G and the resonator is in the coherent state $c|\sqrt{\bar{n}}) = \sqrt{\bar{n}} |\sqrt{\bar{n}})$ but the frequency of the drive is now tuned to the resonance appropriate to the atom in B; so that: $\omega_D = \omega_R$. Next apply the RWA by looking for a wave function in the form: $\psi = \sum C_{G,n}(t) e^{-in \omega_Dt} |n \rangle$. From \eqref{equation:resonator_hamiltonian}, \eqref{equation:effective_resonator_hamiltonian} one obtains:
\begin{equation}
\label{equation:ode_for_cgn}
\frac{dC_{G,n}}{dt} = in\chi C_{G,n} + \Gamma \left( \sqrt{n} C_{G,n-1} - \sqrt{n+1}C_{G,n+1} \right) - \left( \frac{\kappa n}{2} \right) C_{G,n}.
\end{equation}
The solution is given by \eqref{equation:solution_form}, \eqref{equation:ode_beta}, supplemented with 
\begin{equation}
\label{equation:solutions_for_alpha}
\begin{gathered}
\dot{\alpha} = \left( i \chi  - \frac{\kappa}{2} \right) \alpha + \Gamma, \\
\alpha(t) = \frac{\Gamma}{(\kappa/2) - i\chi} \left( 1 - \exp \left(\left( i\chi - (\kappa/2) \right) t \right) \right) + \alpha(0) \exp \left( \left( i \chi - (\kappa/2) \right) t \right).
\end{gathered}
\end{equation}
The initial value of the coherent state $\alpha(0)$ [here equal to $\sqrt{\bar{n}}$] evolves into a lower value $\gamma_L = \Gamma/ ((\kappa/2) - i\chi)$ when the driving frequency is detuned by  $\chi$ from the resonance when the transmon is in G. During this evolution \eqref{equation:probability_of_not_next_jump} applies, so the probability of a jump prior to reaching the new stationary state can be large.
\par
The driven damped oscillator reaches a steady state where the average occupation is $\bar{n}$, when the atom is in G, and the cavity driven at resonance. This result is contained in \eqref{equation:solution_beta}, \eqref{equation:ode_beta} which yield $\alpha \to \sqrt{\bar{n}}$; $t \to \infty$. Suppose now that one wishes to make a measurement of a quantum transition relative to this state. In this case a null measurement occurs when the resonator remains in the coherent state $|\gamma )$ where $c|\gamma ) = \gamma |\gamma )$. The observation of a transition from the coherent steady state $|G \rangle | \sqrt{\bar{n}} )$ is described by the shifted operator:
\begin{equation}
\mathcal{I}_{\gamma} \rho = i \kappa (c - \gamma) \rho (c^\dagger - \gamma^*)  = i\kappa b \rho b^\dagger,
\end{equation}
where '$b$' is the annihilation operator relative to the coherent state $\gamma$. In this case conservation of probability leads to:
\begin{equation}
\label{equation:effective_hamiltonian}
H_{eff} = H_R - i \hbar \left( \frac{\kappa}{2} \right) c^\dagger c - i \hbar \kappa \left( \frac{\gamma \gamma^*}{2} \right) + i\hbar \kappa \gamma^* c.
\end{equation}
If the atom is in G and $\omega_D = \omega_C$ the effective Hamiltonian in the RWA as a function of $b$ is: 
\begin{equation}
\frac{H_{eff}}{\hbar} = -i \left( \frac{\kappa}{2} \right) b^\dagger b + ib \left( \Gamma - \left( \frac{\kappa \gamma^*}{2} \right) \right) - ib^\dagger \left( \Gamma - \kappa \gamma^* + \left( \frac{\kappa \gamma}{2} \right) \right) + i\Gamma(\gamma - \gamma^*).
\end{equation}
Noting that $\Gamma = \kappa \sqrt{\bar{n}}/2$ one finds that when the system is in the coherent state of the original basis [$\gamma = \sqrt{\bar{n}}$] the effective Hamiltonian in the shifted basis is:
\begin{equation}
\frac{H_{eff}}{\hbar} = -i \left( \frac{\kappa}{2} \right) b^\dagger b.
\end{equation}
So if the driven resonator/transmon system starts out in the state G, $b = 0$ it stays in that state. Even though the Hamiltonian drive is balanced by damping there is no quantum jump unless an additional external field drives a Rabi flop or if $\omega_D$ is changed. The vacuum state $| 0 \rangle$ with eigenvalue $b =0 $ of the shifted basis corresponds to the coherent state $\gamma = \sqrt{\bar{n}}$ of the original basis. In this sense the steady state of the classical dissipative system becomes a quantum ground state in the next jump formalism. 

\subsection{Comparison of next jump theory with the stochastic Schr\"odinger equation for heterodyne detection}\label{comp}
The state of the driven cavity gives information about the state of the qubit due to their dispersive coupling. Above we have discussed how the time of its next quantum jump is sensitive to the state of the cavity. This can be seen from the differing solutions to \eqref{equation:ode_for_cgn} and \eqref{equation:ode2_for_cgn}. In most current experimental arrangements  \cite{devoret} the state of the cavity and therefore that of the qubit, is determined by the readout of a heterodyne current. This is the current generated on a microwave detector which is due to the combination of the cavity field with the field of a much stronger beam. 
\par
The heterodyning beam has a different frequency  $\omega_h = \omega_c + \Delta \omega$ than the cavity. The heterodyne beam impinges on the detector directly, without interacting with the cavity.  The total field to hit the detector is in the classical limit
\begin{equation}
E = \left( \frac{B}{\sqrt{2}} \right) \exp i(\omega_c + \Delta \omega)t + \left( \frac{F}{\sqrt{2}} \right) \exp i\omega_ct + cc,
\end{equation}
where $B$, $F$ are the effective field of the heterodyne beam and the cavity emission. The current recorded by the detector includes a DC component plus a term at the difference frequency which is the heterodyne current:
\begin{equation}
E^2 = |B|^2 + F^2 + I_h ,
\end{equation}
\begin{equation}
\label{equation:interference_term}
I_h = \left( \frac{BF^*}{2} \right) \exp (i \Delta \omega t) + cc.
\end{equation}
In the steady state the cavity emits $\kappa \bar{n}$ photons per second; so in the steady state $F = \sqrt{\kappa \bar{n}}$. Even though we shall take $|B|^2 \gg |F|^2$ the fluctuations in the externally imposed beam must be included. In a time $\Delta t$ the number of photons in the external beam is: $n = \langle \langle n \rangle \rangle + \Delta \zeta$ where $ \langle \langle n \rangle \rangle = |B|^2 \Delta t$. Measurement of the heterodyne current at frequency $\Delta \omega / 2 \pi$ is conditioned on measurement of a current equal to $\langle \langle n \rangle \rangle/ \Delta t$ in the DC channel of the photodetector’s current. The fluctuating signal represented by $\Delta \zeta$ provides the source of photons for the heterodyne current. In order that the shot noise in this interval of time be given by $\sqrt{\langle \langle n \rangle \rangle}$ we take $\Delta \zeta$ as being Gaussian distributed so that in the continuous limit 
\begin{equation}
\label{equation:distribution}
\langle \dot{\zeta}(t) \dot{\zeta}(t') \rangle = B^2 \delta(t-t').
\end{equation}
The fluctuation in number is equivalent to a fluctuation in the field of the imposed beam of  $\Delta B(t)  = \dot{\zeta}(t) /B$. The inequality $|B|^2 \gg |F|^2$ means that $\Delta t$ is small compared to the time over which the cavity response is changing.
\par
The probability of recording a given trajectory $\zeta(t)$ is determined by the stochastic Schr\"odinger Equation which we now motivate
heuristically. The deductive presentation following reference \cite{wiseman, wiseman-milburn} is in section~\ref{meas}. According to \eqref{equation:interference_term} the measurement operator for a heterodyne current includes the product of the cavity and external fields. The cavity field is due to the quantum processes in the cavity and is given by $F \to \sqrt{\kappa}c$. The heterodyning field is treated classically and is given by $\dot{\zeta}/B$. In this way observation of a heterodyne emission is represented in the continuum limit by:
\begin{equation}
\label{equation:kraus_operator}
\mathcal{I}_h \rho = -i \sqrt{\kappa} \left( c \rho \left( \frac{\dot{\zeta}}{B} \right) e^{-i \Delta \omega t} +  \left( \frac{\dot{\zeta}}{B} \right)  e^{i \Delta \omega t} \rho c^\dagger \right).
\end{equation}
The additional terms which put the operator \eqref{equation:kraus_operator} into the form: $\mathcal{I}_h \rho = O \rho O^\dagger$ vanish in the continuum limit as shown in section~\ref{meas}. This leads to the effective Hamiltonian:
\begin{equation}
H_{eff} = H_R - i \hbar \left( \frac{\kappa}{2} \right) c^\dagger c + i \hbar \left( \frac{\sqrt{\kappa} \dot{\zeta}}{B} \right) (\exp(-i \Delta \omega t))c .
\end{equation}
The probability of recording a given trajectory $\dot{\zeta}(t)$ is now given by :
\begin{equation}
i \dot{\psi} = \left(-i \Gamma (c - c^\dagger) - i \left( \frac{\kappa}{2} \right) c^\dagger c + i \left( \frac{\sqrt{\kappa} \dot{\zeta}}{B} \right)  e^{-i \phi(t)} c \right) \psi ,
\end{equation}
where we have taken $\omega_c =0$; $\Delta \omega t = \phi(t)$. A solution can again be found in the form \eqref{equation:solution_form} with
\begin{equation}
\label{equation:combined_odes_for_alpha_beta}
\begin{gathered}
\alpha = \frac{2 \Gamma}{\kappa} + \left( \alpha(0) - \frac{2 \Gamma}{\kappa} \right) e^{- \kappa t/2}, \\
\dot{\beta} = -\Gamma \alpha + \alpha \left( \frac{\sqrt{\kappa}}{B} \right) \dot{\zeta} e^{-i \phi(t)}.
\end{gathered}
\end{equation}
The initial value $\alpha(0)$ evolves to the steady state value $\gamma = \sqrt{\bar{n}}$, which value will be chosen for this example. In this limit:
\begin{equation}
\label{equation:beta_solution2}
\beta = -\Gamma \sqrt{\bar{n}} t + \sqrt{\bar{n}} \left( \frac{\sqrt{\kappa}}{B} \right) T(t) + \beta(0),
\end{equation}
where $T(t) = \int \limits_{0}^t ds \dot{\zeta}(s) e^{-i\phi(s)}$ and now $\beta (0) = -\bar{n}/2$. From \eqref{equation:distribution} $\langle \langle T(t) T^*(t) \rangle \rangle = B^2 t$. Therefore, $T$ is a complex random variable that is Gaussian distributed as $P(T) = (1/\pi B^2 t)\exp(-TT^*/B^2 t)$; and defining the average current $I = T/t$; its distribution as a complex variable is:
\begin{equation}
 P(I) = \frac{t}{\pi B^2 } \exp(-t I^2/B^2).
\end{equation}
\par
Although a given current trajectory has a Gaussian distribution the recorded current also involves the probability of making that measurement. The combination of these probabilities completes implementation of the SSE. In this case \eqref{equation:infinite_sum}; \eqref{equation:beta_solution2} yield the combined probability distribution for $I$ as being 
\begin{equation}
\label{equation:probability_current}
P_h(I) = P(I) || \psi(I) ||^2 = \left( \frac{t}{\pi B^2} \right) \exp \left( -t \left( \frac{I}{B} - \sqrt{\kappa \bar{n}} \right)^2  \right).
\end{equation}
So for large times the probability is sharply peaked at $I = B\sqrt{\kappa \bar{n}}$ which is the classical limit of the heterodyne current. To determine $\bar{n}$ with a fractional accuracy $\delta$ requires that the wave function be narrowed down for a time $\kappa t > 1/\delta^2 \bar{n}$. If the measurement process starts from the state with  $\alpha = 0$ then an additional time scale $t \sim 1/\kappa$ enters the measurement which is determined by the evolution of $\alpha$ according to \eqref{equation:combined_odes_for_alpha_beta}.
\par
When the measurement has been sufficiently long the maximum probability of \eqref{equation:probability_current} is attained: $I(t) = \dot{\zeta} e^{-i\phi} = B\sqrt{\kappa \bar{n}}$. Making this substitution into the SSE leads to \eqref{equation:effective_hamiltonian} except for a normalization factor. So when the measurement has been carried out longer than $1/ \kappa$ the heterodyne protocol is the same as the evolution according to the effective Hamiltonian \eqref{equation:effective_hamiltonian} which describes the null evolution between jumps.

\subsection{Short time measurements of the state of the qubit}\label{2.3}
Although the heterodyne protocol as dictated by the SSE converges to the null measurement evolution at long times their fidelity/noise characteristics differ at short times. In the limit of large $\bar{n}$; $|\chi|$, an efficient  measurement of the next jump has advantages. 
\par
\begin{figure}[h]
\begin{center}
\epsfig{file=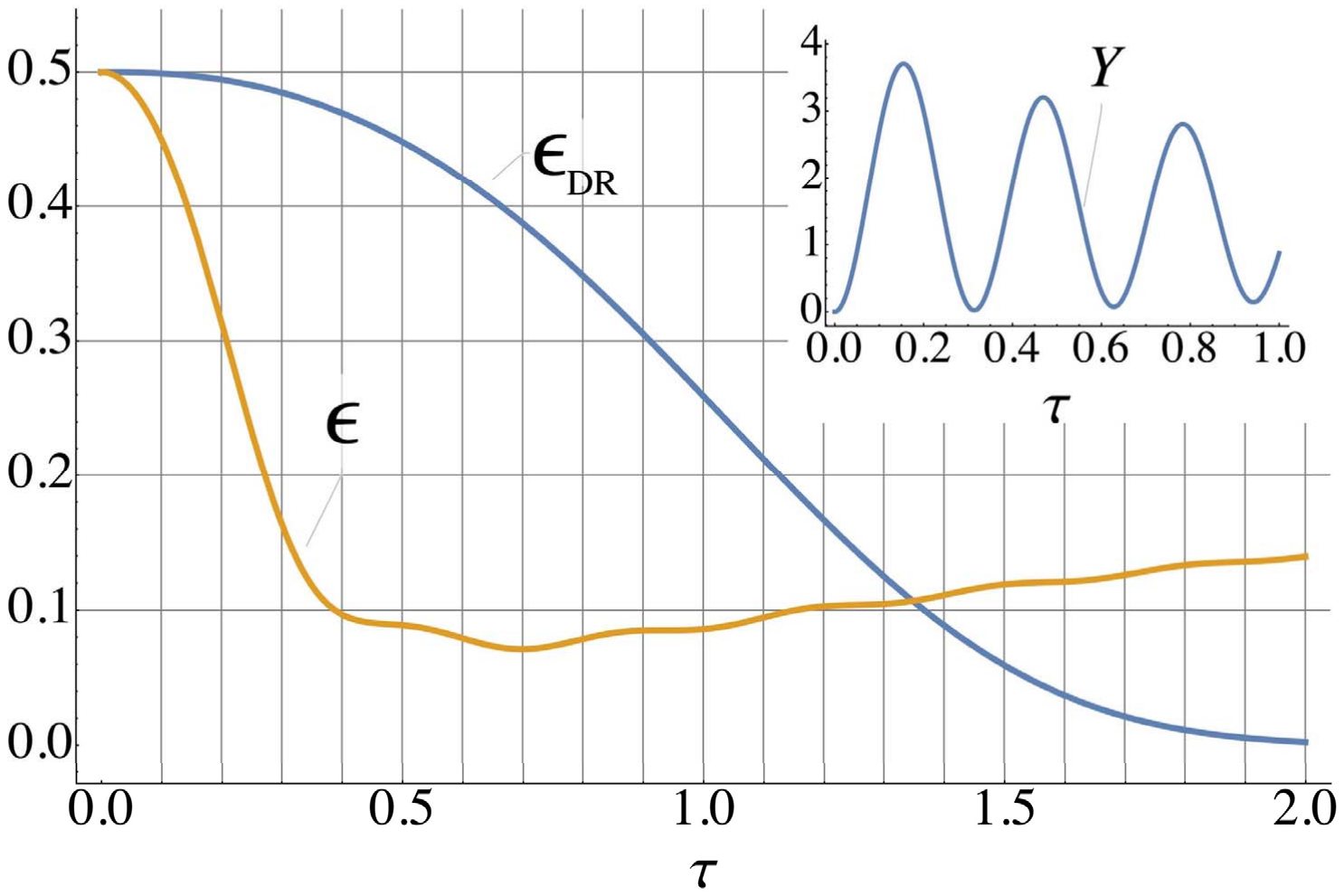, height=3.5in, width=6in}
\end{center}
\caption{Comparison of error for qubit state detection for dispersive homodyne readout $\epsilon_{DR}$
and next jump readout $\epsilon$. The dimensionless time is $\tau=\kappa t$. The error for the next jump detection decreases with increasing dispersion and for this curve we have taken $\chi/\kappa=20$; for dispersive readout $\chi/\kappa=1/2$ which optimizes the SNR. In each case the resonator drive is set to $\bar{n}=100$. For dispersive homodyne readout
the error becomes very small at times longer than the resonator lifetime. For coherent qubit detection via the next jump the error is minimized at short time and becomes $50\%$ at long times [not shown]. However, at very short time $\chi t<1$ the error
 $\epsilon$ is large as dispersion prevents one from determining whether the atom is in $|G\rangle$ or$ |B\rangle$. 
Inset: For large dispersion the time dependence of the arrival of the
next photon has large oscillations. The scaled logarithmic decrement Y of the norm as a function of time is plotted in the inset for 
$\chi/\kappa=20$.}
\label{figure}
\end{figure}

For the two-level atom dispersively coupled to a resonator a measurement of the next jump of the resonant cavity yields information on the state of the qubit. If the qubit is in $|B \rangle$ and the drive frequency is on resonance for this state
then the expected time to the next jump is $\bar{t}_j \sim (3/\kappa \Gamma^2)^{1/3}$. For sufficiently large $\bar{n}$ this time is shorter than $1/ \kappa$. If instead of being in $|B \rangle$ the qubit was in state $|G \rangle$ when the drive is turned on there is also a chance of recording a jump. To evaluate this error rate we introduce the probabilities $P_G(t)$; $P_B(t)$ that the qubit is in state G; B and there has been a click in the detector within time "$t$" of the drive
being turned on: $P_G(t) = 1- W(\chi, t)$ ; $P_B(t) = 1 - W(0,t)$ where $W(\chi, t)$; $W(0,t)$ are given by \eqref{equation:ode_beta}, \eqref{equation:infinite_sum}, \eqref{equation:solutions_for_alpha}. Using the condition that a click has been observed at or before a time '$t$' the error rate is: $\epsilon = P_G/(P_G + P_B)$. 
Homodyne and heterodyne techniques have different errors at short time. First we compare the error for next photon readout to the error for dispersive readout using the homodyne technique. In this case the SNR at short times follows from the application 
of~\eqref{equation:probability_current} or by use of the quantum Langevin equation~\cite{didier} to get
\beq
SNR = \frac{1}{\sqrt{18}}\frac{\Gamma}{\kappa} (\kappa t)^{5/2}.
\eeq{SNR-homod}
A comparison of these metrics is made possible via the relation between the error $\epsilon_{DR}$ for homodyne dispersive readout and its SNR \cite{krantz}:
\begin{equation}
\epsilon_{DR} = \frac{1}{2} \mathrm{erfc} \left( \frac{SNR}{2} \right).
\end{equation}
Figure \ref{figure}~\cite{PPprl}
displays a comparison of the error for dispersive homodyne readout compared to the next jump. The inset contains a plot of the logarithmic decrement of the time-wise distribution of the next jump:
\begin{equation}
Y = - \frac{1 + (2 \chi / \kappa)^2}{\bar{n} W} \frac{dW}{d\tau}.
\end{equation}
Information in the oscillations at short time may allow for a smaller error in the proposed measurements. 
Whereas a click is interpreted as occupation of $|B\rangle$  the absence of a click in $[0,t]$ can be interpreted as occupation of $|G\rangle$. The error $\epsilon_0$  in this application of a null measurement is due to the possibility that the atom does not emit a 
photon while in $|B\rangle$: $\epsilon_0 =W(0,t)/[W(0,t)+W(\chi,t)]$. For the parameters used for Fig.~\ref{figure}
$[\chi/\kappa = 20; \; \bar{n} = 100 ]$
$\epsilon_0 \sim .3$; $\epsilon \sim .1$, for $\kappa t = .5$. When $\kappa t = .7$ the error of readout due to a null
measurement has dropped to $\epsilon \sim  .1$.

In subsection~\ref{homodyne} we will derive
  eq.~\eqref{SNR-homod} using the formalism developed in section~\ref{meas}.
  We will also show that the time dependence $SNR\propto (\kappa t)^{5/2}$ is
  peculiar to the homodyne scheme and that in the heterodyne scheme it is
  instead $SNR\propto (\kappa t)^{3/2}$.
 These signal to noise ratios are optimized with respect to the dispersion. In the case of homodyne readout this occurs for 
 $\chi=\kappa/2$ and heterodyne readout is optimized for $\chi \rightarrow \infty$. For next photon detection the short time error decreases to zero as dispersion increases. To see this consider the norm $W(\chi,t)$ at short times $\kappa t <1$ which follows from 
 eqs.~(\ref{equation:ode_beta},\ref{equation:solutions_for_alpha}):
$W(\chi,t)\sim \exp\left[-{\kappa^3\bar{n}\over 2\chi^2}  \left(t-{\sin \chi t \over \chi}\right)\right]$. 
During the time interval $1/\chi <t\ll 1/\kappa$,
$W(\chi,t)\sim \exp\left[-\kappa^3\bar{n}t/2\chi^2\right]$.Taking for example $\kappa t \sim [12/\bar{n}]^{1/3}$ which is the expected 
time to the next jump, then $W ( \chi , \bar{t}_j) \sim \exp\left[-(3 / 2)^{1/3} (\kappa \bar{n}^{1/3} / \chi )^2 \right]$ and for large $\chi$,
$P_G \sim (\kappa \bar{n}^{1/3}/\chi)^2 \sim \epsilon$.
\subsection{Lifetime of 2 level system coupled to a driven resonator}\label{2.4}
The upper level of a 2 level quantum system acquires a finite lifetime due to its coupling to a driven resonator. This lifetime $1/ \beta_B = \sqrt{\pi}/2 \sqrt{2}\Gamma$ exists even in the limit $\kappa \to 0$. It is similar to radiation damping except here the flow of energy is to higher excited levels $|n\rangle$. This term plays a key role when a transmon is driven by Rabi flopping between B and G. It sets the fastest time scale for the system. 
\par
We start with the 2 level system coupled to a resonator. Now the drive is set to be resonant with the atom in B. When externally driven transitions are included the equations become:
\begin{equation}
\frac{dC_{G,n}}{dt} = i \Omega^* C_{B,n} + \left( \frac{dC_{G,n}}{dt} \right)_R,
\end{equation}
\begin{equation}
\frac{dC_{B,n}}{dt} = i\Omega C_{G,n} + \Gamma \left( \sqrt{n} C_{B,n-1} - \sqrt{n+1} C_{B, n+1} \right) - \left( \frac{\kappa n}{2} \right) C_{B,n},
\end{equation}
where $(dC_{G,n}/dt)_R$ is given by the rhs of \eqref{equation:ode_for_cgn} and $\Omega$ is the Rabi flopping frequency. We analyze this process in the limit where: $\Gamma^2/ \chi^2 \ll 1$; so that $|C_{G,2}/C_{G,1}|^2 \ll 1$. Neglecting $C_{G,n}$ for $n>1$ yields: 
\begin{equation}
\label{equation:dcb0dt}   
\frac{dC_{B,0}}{dt} = i \Omega C_{G,0} - \Gamma C_{B,1},
\end{equation}
\begin{equation}
\label{equation:dcg0dt}
\frac{dC_{G,0}}{dt} = i \Omega^* C_{B,0} - \Gamma C_{G,1},
\end{equation}
\begin{equation}
\label{equation:dcg1dt}
\frac{dC_{G,1}}{dt} + \left( \frac{\kappa}{2} - i\chi \right) C_{G,1} = \Gamma C_{G,0}.
\end{equation}
The flow of probability to higher levels leads to a relation which closes the equations~\cite{PPprl} (see section 4 for more details):
\begin{equation}
\label{equation:cb1}
C_{B,1} = \sqrt{\frac{2}{\pi}} C_{B,0}.
\end{equation}
Use of \eqref{equation:cb1} in \eqref{equation:dcb0dt} gives a lifetime $1/\beta_B$ to the upper level $C_{B,0}$ where:
\begin{equation}
\frac{\beta_B}{2} = \sqrt{\frac{2}{\pi}} \Gamma. \label{47}
\end{equation}
The solution to \eqref{equation:dcb0dt}, \eqref{equation:dcg0dt}, \eqref{equation:dcg1dt}, \eqref{equation:cb1} displays motion on two well separated time scales: the short time scale $2 / \beta_B$ and the long time scale $1/\gamma$; where:
\begin{equation}
\gamma = \frac{2 |\Omega|^2}{\beta_B}.\label{lts}
\end{equation}
 If a continuous measurement of the system finds that the next jump has not occurred for a time longer than $2/ \beta_B$ then there is a lull and a warning that the jump will take place on the long time scale $1/\gamma$. During this time the transmon is in a coherently evolving superposition of its B and G  levels.  During this lull there is an opportunity to rotate the wave function amplitudes and affect the future jump statistics \cite{devoret}. \par
In section~\ref{lifetime} we present the theory for the dark periods which occur when an additional level ``D'' is introduced into the transmon system. It is coupled to ``G'' via Rabi flopping. As ``D'' is not coupled directly to the resonator it can have a long lifetime. This theory is an extension of the 3 level atom [section~\ref{nextPhotonEmission}] to include the quantum mechanical treatment of the readout resonator which is coupled to ``G,B''.

\section{A quick review of continuous measurement theory}\label{meas}
The summary of the theory of continuous measurement by photodetectors given in this section relies 
heavily on refs.~\cite{wiseman-milburn,wiseman} (see also \cite{carmichael,milburn,wiseman-milburn-2}). The results described here are well known and the purpose of this section is twofold. 
First of all, we want to make the limitation and domain of applicability of the Stochastic Schr\"odinger 
Equation (SSE) as explicit as possible. 
Specifically, we will show that while the SSE adequately describes the measurement protocols known as 
heterodyne or homodyne detection, it shares with those protocol an intrinsic limitation in time  resolution. 
We will find that both SSE and heterodyne/homodyne detection are inadequate to describe short time scale measurements but we
will also show that {\em other measurement schemes exist, which do not share that limitation.} This is the other 
purpose of this section. 
The description of other measurement schemes, that can give information about the transmon-cavity system on
short time scales, will begin in section~\ref{lind}.

We will consider a system made by an observed subsystem, described by a state vector $\psi$ in a Hilbert space $\mathcal{H}$ and an ancilla, described by a
vector $\prod_a |\alpha_a\rangle$ in a product Hilbert space $\prod_a \mathcal{H}_a$. 
Measurement is described by initializing the ancilla Hilbert space in the vector $\prod_a |0\rangle$ and letting
the system interact during the time interval $[n\epsilon, (n+1)\epsilon]$ with the $\mathcal{H}_{a=n}$ copy of ancilla 
Hilbert space. 
We will make use of a preferred basis of the Hilbert space $\mathcal{H}_a$ which we denote by $|\alpha\rangle$, 
$\alpha=0,...,N$. In the time 
interval $[n\epsilon, (n+1)\epsilon]$, the sytem $\psi$ and the appropriate 
component of the ancilla $|0\rangle$ evolve under a 
Hamiltonian $H$. On the initialized state $\psi \otimes | 0\rangle$  $H$ is
\beq
H \psi \otimes |0\rangle = \sum_\alpha (\tilde{E}_\alpha \psi) |\alpha\rangle, \quad 
H \psi \otimes |\alpha\rangle = (\tilde{E}_\alpha^\dagger \psi) |0\rangle ,
\eeq{qm1}
where the $E_\alpha$ are some operators acting on $\mathcal{H}$.
The evolution from $t=n\epsilon$ to $t=(n+1)\epsilon$ to $O(\epsilon^2)$ is
\beq
\psi \otimes |0\rangle \rightarrow \psi \otimes |0\rangle -i\epsilon \sum_\alpha (\tilde{E}_\alpha \psi) \otimes |\alpha\rangle  
-{1\over 2}\epsilon^2 \sum_\alpha (\tilde{E}_\alpha^\dagger \tilde{E}_\alpha \psi) \otimes | 0\rangle  .
\eeq{qm2}
To describe a continuous measurement one must choose $\tilde{E}_0=O(1)$, $\tilde{E}_\alpha = O(\epsilon^{-1/2})$ so we
define $\tilde{E}_0\equiv E_0$, $\tilde{E}_\alpha =\epsilon^{-1/2} E_\alpha$, $\alpha\neq 0$.
A strong (projective) measurement on the $\mathcal{H}_a$ Hilbert space effects a weak measurement of $\psi$.  
The probability of recording the ancilla in the state $|\alpha\rangle_{a=(n+1)\epsilon}$  is given by the diagonal 
entry $\langle \alpha | \rho_T | \alpha\rangle_{a=(n+1)\epsilon}$ of the density matrix of the system, $\rho_T$. Its time derivative in
the continuum limit is
\bea
\langle 0 | \dot\rho_T  | 0 \rangle_t &=& \left(-iE_0 -{1\over 2} \sum_{\alpha\neq 0} 
E^\dagger_\alpha E_\alpha \right) \langle 0 | \rho_T | 0 \rangle_t  + \langle 0 | \rho_T | 0 \rangle_t
 \left(iE_0 -{1\over 2} \sum_{\alpha\neq 0} E^\dagger_\alpha E_\alpha \right)  , \label{qm3a} \\
 \langle \alpha |  \dot\rho_T  | \alpha \rangle_t &=&   E_\alpha  \langle 0 | \rho_T | 0 \rangle_t
  E^\dagger_\alpha , \qquad \alpha\neq 0   .
 \eea{qm3}
 The subscript $t$ reminds us that there is a {\em different} ancilla Hilbert space at each time $t$.
Since the $n$-th copy of the ancilla space decouples after interacting with the observed subsystem in the time interval $[n\epsilon,(n+1)\epsilon]$  and at $t=(n+1)\epsilon$ the $(n+1)$-th copy of the ancilla state is initialized at $|0\rangle_{a=(n+1)\epsilon}$, 
we see that in the continuum limit at each time the density
matrix is
$ \rho_T = |0\rangle \langle 0 |_t \rho_T |0 \rangle \langle 0|_t $. Thanks to this property we can 
 obtain a reduced  evolution equation by tracing the density matrix over the ancilla variables to define the density matrix of the observed
 subsystem
$\rho=\sum_\alpha \langle \alpha | \rho_T | \alpha \rangle_t $, which obeys the evolution equation
\beq
 \dot\rho = \sum_{\alpha\neq 0} E_\alpha \rho E^\dagger_\alpha -iH_{eff} \rho + i\rho H_{eff}^\dagger,
\qquad H_{eff} = E_0 -{i\over 2} \sum_{\alpha\neq 0} E^\dagger_\alpha E_\alpha .
\eeq{qm4}
Notice that $E_0^\dagger=E_0$. In quantum information theory the evolution in equation~\eqref{qm4}  is 
called a (continuous) quantum 
channel; the $E_\alpha$ are called the Kraus operators. 

We will concentrate on the case $\alpha=0,1$ and call $E_1\equiv E$. The evolution equation becomes
\beq
\dot\rho = \mathcal{L}\rho \equiv \mathcal{I}\rho -i \mathcal{L}_0\rho.
\eeq{qm5}
The {\em time dependent} Lindblad operator  $\mathcal{L}$ has been decomposed in Eq.~\eqref{qm5} into the sum of of a ``free'' term
 $\mathcal{L}_0\rho =H_{eff} \rho - \rho H_{eff}^\dagger$, describing evolution between observations, plus the interaction term
$\mathcal{I}\rho = E\rho E^\dagger$, describing observations.  Here we will use perturbation theory in $\mathcal{I}$
to find the evolution of a quantum system, conditioned on the measurement of photocurrents (rather than individual 
photons).
 
To describe a measurement of a photocurrent, we must understand the evolution of the system over a time interval 
$\Delta t \gg \epsilon$, in which a large number of photons $n\gg 1$ is detected.  Photons are detected by a photodetector
which measures a field obtained by superimposing a large classical signal $\beta/\sqrt{\kappa}$ to the photons
coming from the cavity containing the observed subsystem (i.e. the atom, or the qubit). Since photodetection is proportional 
to the field amplitude, the Kraus operator is in this case
\beq
E=\sqrt{\kappa} c + \beta .
\eeq{qm6}
The time step $\Delta t$ must be large enough to allow multiple photodetections but so short that the evolution of 
$E$ under $H_{eff}$ is negligible~\cite{wiseman-milburn}.
By defining $H_{eff}=H'_{eff} -{i\over 2} \beta \beta^*$ we find that the evolution of the reduced matrix element 
{\em conditioned on observing $n$ photons in the time interval $\Delta t$} is 
\bea
&&\rho(t+\Delta t) =  e^{-\beta \beta^*\Delta t} e^{-iH'_{eff}\Delta t} \int_0^{\Delta t} dt_1 \int_0^{t_1} dt_2... 
\int_0^{t_{n-1}} dt_n  \nonumber \\ &&
e^{iH'_{eff}t_1}E e^{-iH'_{eff}t_1}...  e^{iH'_{eff}t_n} E e^{-iH'_{eff}t_n} \rho 
e^{iH'^\dagger_{eff} t_n} E^\dagger e^{-iH'^\dagger_{eff} t_n} ....
e^{iH'^\dagger_{eff} t_1} E^\dagger e^{-iH'^\dagger_{eff} t_1} e^{iH'^\dagger_{eff}\Delta t} \nonumber \\
&& \approx 
{(\Delta t)^n \over n!} e^{-\beta \beta^*\Delta t} \left(1-iH'_{eff}\Delta t - {1\over 2} H'^2_{eff} (\Delta t)^2 
\right) E^n \rho
E^{\dagger \, n} \left( 1+iH'^\dagger_{eff} \Delta t -{1\over 2} H'^{\dagger\, 2}_{eff} (\Delta t)^2  \right). \nonumber \\ 
\eea{qm7}
In this equation we have discarded terms that become negligible in the limit $\Delta t \rightarrow 0$ while also 
rescaling $\beta$. To find out the correct scaling we recall that eq.~\eqref{qm7}  is the evolution
of the density matrix conditioned on observing $n$ photons in the time interval $\Delta t$. To obtain a continuous limit,
this number should be large. Now, eq.~\eqref{qm7} can be written as
\bea
\rho(t+\Delta t) &=&  {(\beta\beta^*\Delta t)^n \over n!} e^{-\beta \beta^*\Delta t} X \rho X^\dagger, \label{qm8} \\
X &=& \left(1-iH'_{eff}\Delta t - {1\over 2} H'^2_{eff} (\Delta t)^2 \right)\left (1 +\sqrt{\kappa} {c\over \beta} \right)^n .
\eea{qm9} 
 The prefactor is a Poissonian distribution in $n$ with mean $\langle\langle n\rangle\rangle =\beta\beta^* \Delta t$.  
 To keep $\langle\langle n\rangle\rangle $ large in the limit
 $\Delta t \rightarrow 0$ we must have $\beta\sqrt{\Delta t}\gg 1$.

It is convenient to choose 
\beq
E_0=H + i {\sqrt{\kappa}\over 2}(c^\dagger \beta -c \beta^*),
\eeq{qm10}
with $H$ any Hermitian operator that remains finite in the limit $\Delta t \rightarrow 0$. With this choice eq.~\eqref{qm5} 
describes free evolution in between observations and detection events when the
detected field is shifted by a classical field~\footnote{We will use this decomposition in eqs.~(\ref{mm1},\ref{mm2},\ref{mm3}).}. With this choice 
\beq
H'_{eff}=H +i {\sqrt{\kappa}\over 2}(c^\dagger \beta -c \beta^*) -{i\over 2} (\sqrt{\kappa} c^\dagger + \beta^*)
(\sqrt{\kappa} c + \beta)=H -i\sqrt{\kappa}\beta^*c -i{\kappa \over 2} c^\dagger c .
\eeq{qm11}
A short calculation shows that, up to terms that vanish faster than $\Delta t$ in the $\Delta t \rightarrow 0$ limit
\bea
1-iH'_{eff}\Delta t - {1\over 2} H'^2_{eff} (\Delta t)^2 &=& 1-iH\Delta t -\sqrt{\kappa} \langle\langle n\rangle\rangle  {c \over \beta} -
{\kappa \over 2} c^\dagger c \Delta t + {1\over 2} \kappa \langle\langle n\rangle\rangle ^2 {c^2 \over \beta^2}  , \nonumber \\
\left( 1+\sqrt{\kappa}{c\over \beta} \right)^n &=& \left( 1 + \sqrt{\kappa} n {c\over \beta} + \kappa {n(n-1)\over 2} {c^2\over \beta^2}\right) .
\eea{qm12}
Since $\langle\langle n\rangle\rangle  \gg 1$, we can write $n= \langle\langle n\rangle\rangle  + x$, with $x$ a Gaussian random variable with zero mean and variance 
$\langle\langle n\rangle\rangle = \beta\beta^*\Delta t$. Another short calculation then gives
\beq
X=1- i\left( H -i{\kappa\over 2} c^\dagger c \right) \Delta t + {\sqrt{\kappa} \over \beta} x c 
+{\kappa \over \beta^2} \left({x^2\over 2} -{x\over 2} -{\langle\langle n\rangle\rangle \over 2} \right) c^2.
\eeq{qm13}
The random variable $x/\beta^2 $ is formally of order $|\sqrt{\langle\langle n\rangle\rangle }/\beta^2|$, so the scaling of $\langle\langle n\rangle\rangle $ and $\beta$ in the 
$\Delta t \rightarrow 0$ limit implies $x/\beta^2 \ll \Delta t$ and the evolution equations is
\beq
\rho(t+\Delta t) =  P(x)\left\{ \rho(t) -
i\left[\left( H -i{\kappa\over 2} c^\dagger c \right) \Delta t + {\sqrt{\kappa} \over \beta} x c\right]\rho + 
i\rho\left[\left( H +i{\kappa\over 2} c^\dagger c \right) \Delta t + {\sqrt{\kappa} \over \beta^*} x c^\dagger \right] \right\},
\eeq{qm14}
where $P(x)$ is the probability distribution of $x$.
To obtain a continuum limit we define the variable $\Delta \zeta \equiv B x /|\beta|$, which is a Gaussian random 
process with zero mean and variance $\sigma= B^2 \Delta t$, call $\Delta \zeta_l$ the Gaussian variable in the time interval $[l\Delta t, (l+1)\Delta t]$, denote with
 $P(\Delta \zeta_l)=(2\pi\sigma)^{-1/2} \exp(-\Delta\zeta^2_l/2\sigma)$ its probability distribution, and redefine
\beq
\rho(t)=\prod_{l=0}^N P(\Delta \zeta_l)\hat\rho(t),   \qquad t=N\Delta t.
\eeq{qm15}
Here $B^2$ is the classical photon number current that is heterodyned to the signal from the cavity, so it is the same as in
 the introduction: $B^2=\langle\langle n \rangle \rangle /\Delta t$.\footnote{Notice that in this formula as in the introduction 
 $\Delta t$ is a {\em finite, nozero} time interval.}
In the continuum limit $\Delta \zeta$ defines a Gaussian white noise $\zeta(t)$
with covariance $\langle\langle \dot\zeta(t) \dot\zeta(t')\rangle\rangle = B^2\delta(t-t')$. Because $P(\Delta \zeta )\rightarrow \infty$ for any finite-variance 
fluctuation in the limit $\Delta t \rightarrow 0$, the evolution equation for $\hat\rho$ has the continuum limit 
\bea
i\dot{\hat\rho} &=& H[\zeta]\hat\rho- \hat\rho H^\dagger[\zeta],  \nonumber \\
H[\zeta] &=& \left( H -{i\over 2} \kappa c^\dagger c +i {\sqrt{\kappa} \over B} \dot\zeta e^{i\phi} c\right) , \qquad
e^{i\phi}\equiv -i {|\beta|\over \beta},
\eea{qm16}
which can be rewritten in terms of an SSE
\beq
i\dot\psi= H[\zeta]\psi=\left( H -{i\over 2} \kappa c^\dagger c + i{\sqrt{\kappa} \over B} \dot\zeta e^{i\phi} c\right) \psi .
\eeq{qm17}
A few comments are necessary now.
\begin{itemize}
\item The wave function $\psi$ is simply an auxiliary tool. It is not fundamental and in fact it describes \underline{some}
but \underline{not} all measurement of the cavity+atom. To arrive at eq.~\eqref{qm17} we made several 
approximations, most 
notably we coarse-grained in time over intervals $\Delta t$ and superimposed the signal from the cavity with a classical
signal $\beta/\sqrt{\kappa}$ that diverges in the limit $\Delta t\rightarrow 0$. 
Since any physical $\beta$ is finite, the continuum limit can be only an approximation.
\item The time interval $\Delta t$ itself is much larger than the time interval between measurements of the ancilla:
$\Delta t\gg \epsilon$. In fact $\Delta t$ should be long enough
to allow for detection of many photons.
\item when $\phi(t)$ is constant eq.~\eqref{qm17} describes homodyne detection while when 
$\phi(t)=\omega t$ with $\omega \gg \kappa$ it describes heterodyne detection.
\item As we will see in subsection~\ref{het1}, heterodyne 
measurement correlates well with atomic states only when it is 
performed for a time $T\gg 1/\kappa$. So, heterodyne detection cannot resolve small time scales, such as $1/\kappa$ or
$1/\bar{n}\kappa$. 
\item {\em The main focus of this paper is instead to describe other measurements (next photon) that can give
information on the transmon-cavity state on time scales shorter than $1/\kappa$}
\item Once a solution $\hat\rho[\zeta]$ of eq.~\eqref{qm16} is found, the expectation value 
$\langle \langle O(T) \rangle \rangle_S$ of an operator $O(T)$ 
{\em conditioned on having recorded an output $\zeta(t)$, $t\in[0,T]$ belonging to a set of trajectories $S$} 
is
\beq
\langle\langle O \rangle\rangle_S=\int_S [d\zeta]P[\zeta]\Tr O(T) \hat\rho(T), 
\qquad P[\zeta]= C \exp\left(-\int_0^Tdt {1\over 2 B^2 } \dot\zeta^2 \right).
\eeq{qm18}
The density $P[\zeta]$ defines a Gaussian white noise stochastic process. The constant $C$ normalizes the 
probability distribution and is necessary to make the functional integral over $S$ finite.
\item By redefining the auxiliary state vector $\psi\rightarrow \exp(Ac^\dagger +B)\psi$, we can change $H[\zeta]$, so 
the stochastic Hamiltonian is not unique. We will discuss this freedom in appendix~\ref{het2}.
\end{itemize}

\subsection{A detour into heterodyne detection}\label{het1}
In this paper we will describe the evolution of a quantum system under the constraint of null detection. In this subsection we show how to 
relate the null-detection evolution, given by the Lindblad equation, to evolution in the heterodyne detection scheme which is commonly used in experimental settings. 
This subsection and the next have been extensively summarized in the introduction, where a Fock space language was used.
Here we will use instead the coherent-state formalism. The reader uninterested in the details of the formalism can jump to section 3.

The toy example we consider here is a driven cavity with Hamiltonian 
\beq
H=-\Gamma(c-c^\dagger).
\eeq{ham}
A null result is recorded when the cavity is in a coherent state $|\gamma)$ such
that $(c-\gamma)|\gamma)=0$ so the appropriate Kraus operators for this system are~\eqref{qm6} with $\beta=\sqrt{\kappa}\gamma$ and 
$E_0$ as in eq.~\eqref{qm10}. 
The effective Hamiltonian then follows straightforwardly from eq.~\eqref{qm11} and the redefinition made after eq.~\eqref{qm6}
($H_{eff}=H'_{eff} -{i\over 2} \beta \beta^*$)
\beq
H_{eff}= E_0 -{i\kappa\over 2} c^\dagger c + i\kappa\gamma^* (c-\gamma)+{i\kappa\over 2} \gamma^*\gamma = -{i\kappa\over 2} c^\dagger c -\Gamma(c - c^\dagger ) 
-{i\kappa\over 2} \gamma^*\gamma +i\kappa\gamma^* c .
\eeq{mm17}
Only one mode with many levels is detected. 
Notice the term $-i\kappa\gamma\gamma^*/2$.
In the next photon approach a transition between two levels of a system with one mode is detected. 
In heterodyne detection instead, the state of the cavity is measured by recording a current issuing from a photomultiplier 
that measures the amplitude
of a field obtained by superimposing the cavity field $c$ with a classical signal $\beta(t)$. The frequencies of $c$ and 
$\beta$ are different and the magnitude of $\beta$ is much larger than $\sqrt{\kappa}|\gamma|$. 
So the question to answer is: why experiments
that measure the state of the cavity by using heterodyne detection track the coherent state of the cavity? Here we follow closely~\cite{wiseman}. 
We need to show that heterodyne detection collapses the cavity into a coherent state in a 
time $O(1/\kappa)$. The linear, conditioned time evolution equation is given by eq.~\eqref{qm17} (cfr. eq. (41) of~\cite{wiseman} )
\beq
i\dot\psi= \left( i{\sqrt{\kappa}\over B}\dot\zeta e^{-i\phi} c + H -{i\kappa\over 2}c^\dagger c \right)\psi.
\eeq{mm18}
The variable $\zeta$ is a Gaussian white noise with covariance $\langle \langle \dot\zeta(t) \dot\zeta(t')\rangle \rangle = B^2\delta(t-t')$.
By substituting the ansatz 
$\psi=\exp(\alpha c^\dagger +\beta)|0\rangle$ into~\eqref{mm18} we get the equations 
\beq
i\dot\alpha= -{i\kappa\over 2}\alpha +\Gamma, \qquad i\dot\beta = -\Gamma\alpha + i{\sqrt{\kappa}\over B}\dot\zeta e^{-i\phi}\alpha.
\eeq{mm19}
The solution obeying initial condition $\alpha(0)=\bar{\alpha}$ and $\beta(0)=0$ is 
\bea
\alpha(t) &=& 2{\Gamma\over \kappa} \left( 1-e^{-\kappa t/2}\right)  +\bar{\alpha} e^{-\kappa t/2}, \nonumber \\
\beta(t) &=&  -2{\Gamma\over \kappa} \left[\Gamma t -T(t) \right] +2{\Gamma\over \kappa} \left(\bar{\alpha} -2{\Gamma\over \kappa} \right) \left( e^{-\kappa t/2}-1 \right) + \left(\bar{\alpha}-2{\Gamma\over\kappa} \right) S(t),
\eea{mm20}
with
\beq
T(t)={\sqrt{\kappa}\over B}\int_0^t ds \dot\zeta e^{-i\phi} , \qquad S(t)={\sqrt{\kappa}\over B}\int_0^t ds \dot\zeta e^{-i\phi} e^{-\kappa s/2}.
\eeq{mm21}
In the heterodyne approximation $\phi(t)=\omega t$, $\omega \gg \kappa $, the expectation values for $T(t)$ are
\beq
\langle T(t)T(t)\rangle =\int_0^t ds e^{2i\Delta s} \approx 0, \quad \langle T(t)T^*(t)\rangle =\int_0^t ds = \kappa t,
\eeq{mm22}
so $T(t)$ is a complex Gaussian variable with probability distribution $P= (\kappa t\pi)^{-1} \exp (-TT^*/\kappa t)$. 
Let us use the average normalized
current $\mathcal{I}(t)=T(t)/t$ as random variable. Its probability distribution is $P=(t/\kappa\pi) \exp(-t \mathcal{I}\mathcal{I}^*/\kappa)$. 
The random variable $S(t)$ is also Gaussian; for $T\gg 1$  it has  covariance $\langle S(t) S^*(t)\rangle =1$.
We see from the first of eqs.~\eqref{mm20} that no matter which output current has been recorded and irrespective of the initial 
state $\bar\alpha$, the cavity evolves towards the classical coherent state $|2\Gamma/\kappa)$ in a time $O(1/\kappa)$.

Following~\cite{wiseman} we compute next the probability $P(\mathcal{I})(t)$ of recording the outcome $\mathcal{I}$. The norm of the initial state is 
$||\psi(0)||=\exp(\bar\alpha \bar\alpha^*/2)$ so for $\kappa t\gg 1$ we get
\beq
P(\mathcal{I})={t\over \kappa\pi} e^{4\Gamma^2/\kappa^2 + \beta(t) + \beta^*(t) -\bar\alpha \bar\alpha^* 
-t\mathcal{I}\mathcal{I}^*/\kappa}, \qquad 
\beta(t)\approx -2{\Gamma \over \kappa} t \left[\Gamma -\mathcal{I}(t)\right] 
-2{\Gamma\over\kappa} \left(\bar\alpha - 2{\Gamma\over \kappa} \right) +\left(\bar\alpha - 2{\Gamma\over\kappa} \right)S(t).
\eeq{mm23}
We see that for $\kappa t\gg 1$  the probability distribution in $\mathcal{I}$ is sharply peaked at its maximum $\mathcal{I}=2\Gamma $
For any other values of $\mathcal{I}$ the probability is exponentially small $P(\mathcal{I})\propto \exp(-t |\mathcal{I}  
-2\Gamma|^2/\kappa)$.

So, a heterodyne readout is a good tracer of the coherent state. The recorded current is sharply peaked at a single value 
$\mathcal{I}=2\Gamma$
and the field in the cavity evolves towards its stationary value as in any classical driven damped oscillator. The characteristic
time scale for the measurement is the same as the damping scale: $1/\kappa$.

\subsection{Relation to the null-measurement evolution}\label{null}
A ``perfect'' current readout is exactly equal at each time $t$ to the maximum-probability current: 
$(\sqrt{\kappa}/B)\dot\zeta e^{-i\phi}=\mathcal{I}(t)=2\Gamma$. By
setting $(\sqrt{\kappa}/B)\dot\zeta e^{-i\phi}=\mathcal{I}(t)=2\Gamma$ in~\eqref{mm18} we obtain an equation 
almost identical
 to the null measurement evolution equation, $i\dot\varphi=H_{eff}\varphi$, where $H_{eff}$ is given by eq.~\eqref{mm17} with
$\kappa\gamma=2\Gamma$. Eq.~\eqref{mm18} seems to miss the term $ -i \kappa \gamma^*\gamma/2$, but the term reappears when one realizes 
that~\eqref{mm18} describes the true evolution of the system only up to a normalization. The probability of an outcome at time
$t$ conditioned on measuring the current $\mathcal{I}(t)=2\Gamma$ is $\exp(-t\mathcal{I}\mathcal{I}^*/\kappa) || \psi (t) ||^2= || \varphi(t)||^2$. The properly normalized wave 
function $\varphi(t)=\exp(-t\mathcal{I}\mathcal{I}^*/2\kappa)\psi$ obeys exactly the equation $i\dot \varphi = H_{eff} \varphi$.

So, for a ``perfect'' heterodyne measurement, that is when the current $\mathcal{I}(t)$ is always exactly at the most probably value, the 
SSE~\eqref{mm18} plus the definition of the the actual probability of an outcome~\eqref{mm23}, tells us that the system evolves
exactly as during a null measurement period, when no quanta $b=c-2\Gamma$ are detected. In other words, when the Gaussian 
distribution is sharp, that is when the measurement has been carried out for longer than $1/ \kappa$, we can forget about the 
complications of the actual heterodyne measurement protocol and study the evolution of the system with the effective 
Hamiltonian~\eqref{mm17}.
Furthermore, if one post selects from the SSE those events where the current $\mathcal{I}(t)$ was equal to its maximum-likelihood value $2\Gamma$, then the system evolves exactly as during a null measurement period for short times as well.

\subsection{Homodyne and heterodyne optimal measurement at short time}\label{homodyne}
Consider a driven cavity with Hamiltonian 
\beq
H=\Gamma^* c+\Gamma c^\dagger + \chi c^\dagger c.
\eeq{ham-chi}
The SSE is almost identical to eq.~\eqref{mm18}
\beq
i\dot\psi= \left( i{\sqrt{\kappa}\over B}\dot\zeta e^{-i\phi} c + H -{i\kappa\over 2}c^\dagger c \right)\psi.
\eeq{mm18chi}
The variable variable $\zeta$ is the same that we defined after
eq.~\eqref{mm18}: a Gaussian white noise with covariance $\langle \langle \dot\zeta(t) \dot\zeta(t')\rangle \rangle = B^2\delta(t-t')$.
By substituting the ansatz $\psi=\exp(\alpha c^\dagger +\beta)|0\rangle$ into~\eqref{mm18chi} we get the equations 
\beq
i\dot\alpha= \omega\alpha +\Gamma, \quad i\dot\beta = \Gamma^*\alpha + i{\sqrt{\kappa}\over B}\dot\zeta e^{-i\phi}\alpha,
\quad \omega =\chi -i\kappa/2.
\eeq{mm19chi}
The solution obeying initial condition $\alpha(0)=\bar{\alpha}$ and $\beta(0)=0$ is 
\bea
\alpha(t) &=& {\Gamma\over\omega} \left(e^{-i\omega t} -1 \right) + e^{-i\omega t} \bar{\alpha}, \nonumber \\
\beta(t) &=&  i{\sqrt{\kappa}\over B}\int_0^t ds \dot\zeta(s) e^{-i\phi(s)}\alpha(s) + ....,
\eea{mm20chi}
where $....$ denotes terms that do not depend on $\zeta$.
When the cavity at $t=0$ is empty $\bar{\alpha}=0$ so eq.~\eqref{mm20chi} gives in the \underline{homodyne} scheme 
($\phi(s)=\phi=\,$constant)
\beq
\beta(t) =  i{\sqrt{\kappa}\over B}\int_0^t ds \dot\zeta(s) e^{-i\phi} {\Gamma\over\omega} \left(e^{-i\omega t} -1 \right).
\eeq{mmm1chi}

We compute next the probability $P(\mathcal{\zeta})(t)$ of recording the outcome $\mathcal{\zeta}$
\beq
P(\mathcal{\zeta})\propto \exp\left( \beta(t) + \beta^*(t) -{1\over 4B^2} \int_0^t ds \dot{\zeta}^2 \right).
\eeq{mm23chi}
In this equation we neglected normalization terms that are  independent of $\zeta$.
The maximum likelihood of the distribution is at
\beq
\dot{\zeta}/B= 2i\sqrt{\kappa}\int_0^t ds  e^{-i\phi} {\Gamma\over\omega} \left(e^{-i\omega t} -1 \right) + c.c.
\eeq{mmm2chi}

Now, the choice made in~\cite{didier} is the following: when the qubit is in $|0\rangle$, the frequancy of the cavity is $\chi$, when
the qubit is in $|1\rangle$, the frequency is $-\chi$. 
Notice that with this choice the photon occupation number is \underline{the same} in both states. The SNR is not maximized by
making the difference in occupation number as large as possible, but instead by making the phase difference of the output current
as large as possible. Namely, the signal $S$ is  the classical current [eq.~\eqref{mmm2chi}] in state $|0\rangle$ minus the classical
signal in $|1\rangle$
\beq
S=\left|  \left.{\dot{\zeta}\over B}\right|_\chi -  \left.{\dot{\zeta}\over B}\right|_{-\chi} \right| .
\eeq{mmm3chi}
Substituting eq.~\eqref{mmm2chi} into~\eqref{mmm3chi} and using the notation $\omega_\pm= \pm\chi -i\kappa/2$ and $\omega_+=-\omega_-^*$ we get
\beq
S= 2\sqrt{\kappa}\left| \int_0^t ds (ie^{-i\phi}\Gamma -ie^{i\phi}\Gamma^*)\left({e^{-i\omega_+ t} -1 \over \omega_+}  + 
{e^{i\omega_+^* t} -1 \over \omega_+^*}\right)  \right|.
\eeq{mmm4chi}
This signal is maximized at any time with the choice $ie^{-i\phi}\Gamma=\pm |\Gamma|$ and for large time it is optimized by 
maximizing 
\beq
{1\over \chi -i\kappa/2} + {1\over \chi +i\kappa/2} = {2\chi \over \chi^2 + \kappa^2/4}
\eeq{mmm5chi}
i.e. by choosing $\chi=\kappa/2$. 
Notice that the signal at times $\kappa t \ll 1$ is proportional to $t^3$ so that the SNR is proportional to $t^3/t^{1/2}=t^{5/2}$.

This is \underline{not} the short term behavior of other dispersive coupling schemes. In particular, in our examples and in~\cite{devoret} 
$\omega_+=-i\kappa/2$, $\omega_-=\chi -i\kappa/2$ with $\chi \gg \kappa$. Moreover, in the heterodyne scheme and in the RWA for $\phi(s)$ the current
is proportional to
\beq
I= \left| \int_0^t ds e^{-i\phi(s)} {\dot{\zeta}\over B} \right| = 
\left| \int_0^t ds {\Gamma^*\over\omega^*} \left(e^{+i\omega^* t} -1 \right) \right| .
\eeq{mmm6chi}
For $\chi \gg \kappa/2$ the current off resonance is negligible, so the signal is equal to the current at resonance $\omega=-i\kappa/2$.
Hence
\beq
S\approx \left| \int_0^t ds {2\Gamma^*\over\kappa} \left(e^{-\kappa t/2} -1 \right) \right| .
\eeq{mmm7chi}
For $\kappa t\ll 1$  $S\propto t^2$ so the SNR is proportional to $t^{3/2}$ instead of $t^{5/2}$.
 
\section{Lindblad operator description of a transmon-cavity  quantum telegraph}\label{lind}
The evolution of a three-level atom in infinite space was studied in ref.~\cite{pp}. Resets of the wave function occur as the result of the detection of a
photon that is scattered off axis from the exciting laser.
There are an infinity of directions of propagation for the scattered photons.
A transmon in a cavity is detected in a
rather different manner. Instead of infinitely many radiation  modes only one mode with many levels is detected; the detection is effected
(indirectly, through heterodyne detection) by recording the state of the cavity, which changes because of the dispersive
coupling of the cavity mode etc. So, to understand this system 
we have to go back to the basic of measurement theory, i.e. to eq.~\eqref{qm4} 
and adapt it to the transmon-cavity system.  An atomic 3-level V-shape system, with one state, D, weakly coupled to the
vacuum G and another state, B, strongly coupled to it gives rise to the ``quantum telegraph'' phenomenon explained in terms
of null detection in~\cite{pp}. To recover a similar behavior in the transmon-cavity case {\em and to relate its parameters to
properties of the cavity} is a very nontrivial check of the correctness of optical measurement theory as well as the
aim of this section.

 The Lindblad operator appropriate to 
the transmon-cavity system is given in eq.~(30) of ref.~\cite{devoret}
\bea
i{d\rho \over dt} &=& \mathcal{L}\rho \equiv
[H_A+H_R,\rho] +i{\kappa\over 2} \big(2c\rho c^\dagger- c^\dagger c \rho - \rho c^\dagger c\big), \nonumber \\
H_A &=&     \Omega_B(t)  |B\rangle \langle G |  + \Omega_B^*(t) |G\rangle \langle B|  + 
\Omega_D\big( |D \rangle \langle G| +  |G\rangle \langle D| \big)        \nonumber \\
H_R &=& \chi  \big( |B\rangle \langle B | -1 \big) c^\dagger c + {\kappa \over 2 i} \sqrt{\bar{n}} \big( c -c^\dagger \big).
\eea{mm1}
Compared with ref.~\cite{devoret} we changed notations as follows: $\hbar=1$, $\Omega_B(t)=i\Omega_{BD}(t)/2$,  
$\Omega_D=i\Omega_{DG}/2$; we also set $\Delta_R=\chi_B=\chi$, $\chi_D=0$. 

The Lindblad operator decomposes into an effective non-Hermitian Hamiltonian --describing coherent evolution in between
 observations-- and a term that describes observation. When the latter is $i\kappa c\rho c^\dagger$,  
 $\kappa \tr c^\dagger c \rho(t) dt$ is
the probability to detect a quantum of the cavity mode $c$ in the time interval $[t, t+dt]$. This would correspond to 
measuring the next quantum jump in a system starting in its true ground state and is calculated in~\cite{PPprl}. 
Here we want something different, we  want to define a null measurement as the detection of the cavity in the 
coherent state 
\beq
c| \gamma ) =\gamma |\gamma ) .
\eeq{mm2}
 So the quanta to be detected must be $b= c- \gamma$. The Lindblad operator now decomposes as $\mathcal{L}\rho=
 H_{eff} \rho - \rho H_{eff}^\dagger + \mathcal{I}\rho$ with
 \bea
 H_{eff}&= & H_A + H_R -i{\kappa\over 2} \gamma \gamma^*  -i{\kappa\over 2} c^\dagger c + i\kappa \gamma^* c , \nonumber \\
 \mathcal{I}\rho &=& i\kappa (c-\gamma) \rho (c^\dagger -\gamma^*).
 \eea{mm3}
 
 Before proceeding further it is necessary to make two comments.
 \begin{itemize}
 \item The first is on the meaning of the parameter $\kappa$. It determines the
 strength of interaction between the cavity mode and the detector that appears in the Kraus operator [see eq.~\eqref{qm6}]. 
 It coincides with the quality factor of the cavity {\em only} in the idealized case that the only way for the photon 
 to escape the
 cavity is by being measured. One can imagine to realize this system with a cavity, perfectly reflecting except for an 
 aperture behind which is a photodetector. The photon may escape through the aperture and be detected, or be reflected back
 into the cavity if not detected. In short: $\kappa$ is the quality factor of the cavity in the limit that 
  side channels through which the photon can escape without being detected are negligible. 
 \item The second observation is that the measurement of the shifted photon $c-\gamma$ is done by a standard 
 photodetector that records a signal made by superimposing the photon escaping from the cavity to a coherent source
 of strength $\gamma$ which is $180^o$ out of phase. In this way an emission from the coherent state generates a null measurement. 
 The procedure is also described by a shifted Kraus 
 operator~\eqref{qm6}, but now with a classical 
  signal that maintains phase coherence with the cavity photon and has a strength which is tuned to whichever the appropriate 
 finite value may be --either $\gamma=\sqrt{\bar{n}}$ or $\gamma=\gamma_L$.
 \end{itemize}
 \noindent
 \hrulefill \\
 The generic form of the Hamiltonian we shall encounter is 
 $H=\omega c^\dagger c + \Gamma c^\dagger + \tilde\Gamma c$. Notice that we did not assume that $\omega$ is real or 
 that $\tilde\Gamma=\Gamma^*$. For any complex $(\omega,\Gamma,\tilde\Gamma)$ the Schr\"odinger equations on a coherent state $\exp(\alpha c^\dagger + \beta)|0\rangle$  is 
\beq
i{d\over dt} e^{\alpha c^\dagger + \beta} |0\rangle = 
\left(i\dot\alpha c^\dagger +i\dot\beta\right) e^{\alpha c^\dagger + \beta} |0\rangle =
\left[ \Gamma c^\dagger + \tilde\Gamma \alpha + \omega \alpha c^\dagger 
\right] e^{\alpha c^\dagger + \beta} |0\rangle .
\eeq{mmm4b}
Equating like terms we get
\bea
\alpha &=& {\Gamma\over\omega} \left(e^{-i\omega t} -1 \right) + e^{-i\omega t} \alpha(0), \nonumber \\
\beta &=& i{\Gamma\tilde\Gamma \over \omega} t +{\Gamma\tilde\Gamma \over \omega^2} \left(e^{-i\omega t}-1 \right)
+ {\tilde\Gamma \over \omega} \left(e^{-i\omega t} -1\right) \alpha(0) +\beta(0).
\eea{mmm4c}
These equations say that a coherent state remains coherent during time evolution.

\subsection{The stationary states of the cavity}\label{stat-stat}
The effective Hamiltonian~\eqref{mm3} at $\Omega=0$ is 
\beq
H_{eff}=\chi(|B\rangle \langle B|-1)(b^\dagger+\gamma^*)(b+\gamma)  -i{\kappa \over 2}\gamma \gamma^* -i{\kappa \over 2}(b^\dagger +\gamma^*)(b+\gamma) 
+i\kappa \gamma^*(b+\gamma) -i{\kappa \over 2}\sqrt{\bar{n}}(b-b^\dagger +\gamma -\gamma^*).
\eeq{mm4}

\subsubsection{Time evolution of the system when the cavity is in the high-field  coherent state}\label{low-state}
When the atom is in $|B\rangle$, the effective Hamiltonian is purely quadratic in $b$ if $\gamma=\sqrt{\bar{n}}$. 
If the atom is not in $|B\rangle$ there is a linear term. As a matrix action on the vector
\beq
\begin{pmatrix} |B\rangle \\ |G\rangle \\ |D\rangle \end{pmatrix},
\eeq{mm5}
the effective Hamiltonian is 
\beq
  H_{eff}  = \begin{pmatrix} -i{\kappa \over 2}b^\dagger b  & 0   \\ 0 & 
 \left(-i{\kappa \over 2} -\chi \right) b^\dagger b -\chi \bar{n} -\chi \sqrt{\bar{n}}(b+b^\dagger) \end{pmatrix} .
\eeq{mm6}
The upper left block is 1 by 1, the lowest right is 2 by 2.
When the atom is in $|B\rangle$ and the cavity is in $|\sqrt{\bar{n}} )$ the system is stable: 
$H_{eff}|B\rangle \otimes|\sqrt{\bar{n}} )=0$. 
Its time evolution is induced by switching on the Rabi frequencies.

If the atom is not in $|B\rangle$, the evolution during a dark period is
\beq
i\dot\psi = \left[\left(-i{\kappa \over 2} -\chi \right) b^\dagger b -\chi \bar{n} -\chi \sqrt{\bar{n}}(b+b^\dagger)\right]\psi .
\eeq{mm7}
The ansatz 
\beq
\psi= \exp\left(\alpha b^\dagger + \beta\right) |0\rangle \otimes (A|G\rangle + B|D\rangle)
\eeq{mm8}
yields the equations
\beq
i\dot\alpha = \left(-i{\kappa \over 2} -\chi \right)\alpha  -\chi \sqrt{\bar{n}}, \qquad
i\dot\beta  = -\chi \sqrt{\bar{n}} \alpha -\chi \bar{n}.
\eeq{mm9}
Here $|0\rangle$ is the vacuum of the $b$ oscillators, defined by $b|0\rangle=0$, so it is the coherent state $|\sqrt{\bar{n}})$. 
The solution with initial condition $\psi |_{t=0}=|0\rangle \otimes (A|G\rangle + B|D\rangle)$ is a particular case of 
eq.~\eqref{mmm4c}
\beq
\alpha=-\chi\sqrt{\bar{n}} {1 - e^{\left(-{\kappa \over 2} +i\chi \right)t} \over i{\kappa \over 2} +\chi},
\qquad
\beta= i\chi \bar{n} {i{\kappa\over 2} t\over i{\kappa \over 2} +\chi} +\chi^2 \bar{n} {e^{\left(-{\kappa \over 2} +i\chi \right)t} -1 \over
\left(i{\kappa \over 2} +\chi\right)^2}.
\eeq{mm10}
In a time of order $t\sim {2/ \kappa} $ the cavity approaches exponentially the coherent state 
\beq
|\sqrt{\bar{n}} - \chi \sqrt{\bar{n}} /(\chi + i\kappa/2) ) =| i\kappa \sqrt{\bar{n}} /(2\chi +i \kappa))\equiv | \gamma_L), \qquad
\gamma_L\equiv i\sqrt{\bar{n}}\kappa/(2\chi+i\kappa) .
\eeq{mm11}
For $\chi \gg 2\kappa$ and $t\simgeq 2/\kappa$ the norm of $\psi$ is $||\psi (t)|| \propto \exp(-\kappa \bar{n} t/2)$. 
In ref.~\cite{devoret} the value is $2\chi /\kappa \approx 3$. Here and elsewhere we denote the scalar product of states 
$\psi,\phi$ with $(\psi,\phi)$ while we use Dirac's notation when states are written as $|\psi\rangle,|\phi\rangle$. 
The norm of $\psi$ is $|| \psi || \equiv \sqrt{(\psi,\psi)}$.
\begin{itemize}
\item
This result and eq.~\eqref{mm11} make sense. They say that when the atom is not in $|B\rangle$, the cavity 
evolves into the low-field state in a time $O(2/ \kappa)$ and that the probability $P$ of \underline{not} detecting the shift 
for a time $t \simgeq 2/\kappa$ is exponentially small, $P=||\psi||^2\propto \exp(-\kappa \bar{n} t)$. 
\item
Notice the shift in the transition energies $-\chi \bar{n}$, which is due to the nonlinear dispersive coupling of the atom to 
the cavity.
\end{itemize}
\subsubsection{Time evolution of the system when the cavity is in the low-field coherent state}\label{high-state}
Equation~\eqref{mm11} shows that if the atom is not in $|B\rangle$, the cavity is quickly driven to the coherent state
$|\gamma_L)$. We can study the evolution of the system when the cavity is in $|\gamma_L)$ at $t=0$. In this case the 
Hamiltonian written in terms of the oscillators $a=c-\gamma_L$ 
is purely quadratic on any state of the form $A|G\rangle + B|D\rangle$, while it acquires terms linear in $a,a^\dagger$ 
if the atom is in $|B\rangle$.
A brief calculation analogous to that leading to~\eqref{mm6} shows that now the effective Hamiltonian is
\beq
  H_{eff}  = \begin{pmatrix} -i{\kappa \over 2}a^\dagger a +i{\kappa\over 2} \left[ (\gamma_L^* -\sqrt{\bar{n}})a 
  -(\gamma_L -\sqrt{\bar{n}} )a^\dagger \right] +2\chi\gamma_L\gamma_L^* & 0   \\ 0 & 
 \left(-i{\kappa \over 2} -\chi \right) a^\dagger a +\chi\gamma_L \gamma_L^*\end{pmatrix} .
\eeq{mm12}
The dispersive shift in frequency for the atomic states $|G\rangle,|D\rangle$ is now $\chi \gamma_L\gamma_L^* \ll \chi \bar{n}$ (when $2\chi \gg \kappa$). 

Time evolution is trivial if the atom is in an arbitrary linear superposition of $|G\rangle$ 
and $|D\rangle$, while if the atom
is in $| B\rangle$, it is again a particular case of eqs.~(\ref{mmm4b},\ref{mmm4c}), so it is solved by the ansatz 
\beq
\psi= \exp\left(\alpha a^\dagger + \beta\right) |0\rangle \otimes |B\rangle ,
\eeq{mm13}
where now $|0\rangle\equiv |\gamma_L)$. 
The equations for $\alpha$ and $\beta$ are now
\beq
i\dot\alpha = -i{\kappa \over 2} \alpha-i{\kappa \over 2} (\gamma_L- \sqrt{\bar{n}}), \qquad
i\dot\beta = i{\kappa \over 2}(\gamma_L^*- \sqrt{\bar{n}}) \alpha +\kappa \sqrt{\bar{n}} \Im \gamma_L.
\eeq{mm14}
The solution with initial condition $\psi|_{t=0}=|0\rangle \otimes |B\rangle$ is
\beq
\alpha=(\sqrt{\bar{n}} -\gamma_L) \left( 1 - e^{-{\kappa \over 2}t} \right)
\qquad
\beta= -i\kappa \sqrt{\bar{n}} \Im \gamma_L t - {\kappa \over 2} |\gamma_L- \sqrt{\bar{n}}|^2  
\left[ t +{2\over \kappa} \left( e^{-\kappa t/2} -1\right)\right] .
\eeq{mm15}
If the atom is in $|B\rangle$, the cavity collapses to the high-field coherent state $|\sqrt{\bar{n}})$ in a time $O(2/\kappa)$.
The probability of a null detection of duration $t \simgeq 1/\kappa$ is proportional to 
$\exp(\beta +\beta^*)\propto \exp(-\kappa  |\gamma_L- \sqrt{\bar{n}}|^2t) \sim \exp(-\kappa\bar{n}t)$ 
when $2\chi \gg \kappa$ (i.e. $|\gamma_L|\ll \sqrt{\bar{n}}$).

In the limit $\kappa \rightarrow 0$, $\kappa\sqrt{\bar{n}}=C$, $C=$constant, we get
$\alpha= Ct/2$, $\beta= -iC\Im \gamma_L t -C^2 t^2/8$ and the norm of the state
at time $t$ is $||\psi||= \exp(\alpha^*\alpha/2 +\beta/2 +\beta^*/2)= 1$. This is
consistent with the absence of dissipation in the limit $\kappa\rightarrow 0$.
In this limit the state diffuses but its norm is constant. The overlap with
the initial state still goes to zero because of diffusion
\beq
|( \psi(0),\psi(t)) |= e^{-C^2t^2/8}.
\eeq{mm15a}
This is the origin of the effective damping of the upper transmon level given by eq.~\eqref{47}.

\subsection{The lifetimes of dark periods}\label{lifetime}
Consider now the case where the drive frequency is chosen so that the readout cavity is resonant when the atom is in $|B\rangle$ and the initial state is chosen to be a superposition of $|G\rangle$ and $|D\rangle$. The system is ``dark'' in that no heterodyne current is observed. If now fields which induce Rabi flopping between $|D\rangle-|G\rangle$ and $|B\rangle-|G\rangle$ are turned on the dark period will eventually terminate. We now compute the lifetime of the dark period with an appropriate version of time-independent perturbation theory. For the intrinsic spontaneous decay rate of $|D\rangle$ we take $\beta_D=0$. During the dark period the system is in the low field
coherent state with $\gamma=\gamma_L$ and one is measuring a jump relative to this coherent state. Therefore the starting point is the effective Hamiltonian~\eqref{mm3} relative to the shifted state which is:
\beq
  H_{eff}  = \begin{pmatrix} H_B +2\chi\gamma_L\gamma_L^* & \Omega_B(t) & 0   \\ \Omega_B^*(t) & 
  H_{\not{B}}+\chi\gamma_L \gamma_L^* & \Omega_D \\ 
 0 &  \Omega_D &  H_{\not{B}} +\chi\gamma_L \gamma_L^* \end{pmatrix} ,
\eeq{mm24}
with 
\beq
H_B\equiv -i{\kappa \over 2}a^\dagger a +i{\kappa\over 2} \left[ (\gamma_L^* -\sqrt{\bar{n}})a 
  -(\gamma_L -\sqrt{\bar{n}} )a^\dagger \right], \qquad H_{\not{B}}= \left(-i{\kappa \over 2} -\chi \right) a^\dagger a .
\eeq{mm25}
This effective Hamiltonian is an extension of
eqs.~(\ref{equation:dcb0dt}-\ref{47})
to include an extra
level $|D\rangle$  and the initial state $\gamma=\gamma_L$.
With the change of basis $|B\rangle \rightarrow \exp(-i 2\chi\gamma_L\gamma_L^* t) |B\rangle$, 
$|G\rangle \rightarrow \exp(-i \chi\gamma_L\gamma_L^* t)|G\rangle$, \\
$|D\rangle \rightarrow \exp(-i \chi\gamma_L\gamma_L^* t)|D\rangle$ and using the RWA we recast 
Hamiltonian~\eqref{mm25} into
\beq
  H_{eff}  = \begin{pmatrix} H_B & \bar\Omega_B & 0   \\ \bar\Omega_B^*& 
  H_{\not{B}} & \Omega_D \\ 
 0 &  \Omega_D &  H_{\not{B}}  \end{pmatrix} , \qquad \bar\Omega_B\equiv \lim_{T\rightarrow \infty} {1\over T}\int_0^T dt 
 \Omega_B(t)  e^{i\chi\gamma_L\gamma_L^* t}.
\eeq{mm26}
Solving for $|B\rangle$ in terms of $|G\rangle$ we get a reduced matrix
\beq
H_{red}-E = \begin{pmatrix} \bar\Omega^*_B (E- H_B)^{-1} \bar\Omega_B + H_{\not{B}} - E & \Omega_D \\
\Omega_D & H_{\not{B}} -E \end{pmatrix} .
\eeq{mm27}
The next step is to expand around the unperturbed solution $H_{\not{B}} \psi=0$. The unperturbed eigenvectors are
\beq 
v_1= \begin{pmatrix} |0\rangle \\ 0 \end{pmatrix}, \qquad v_2=\begin{pmatrix} 0 \\ |0\rangle \end{pmatrix}, \mbox{ with } a|0\rangle=0.
\eeq{mm28}
The equation for $E$ is given by $Ev = \langle 0 | H_{red} | 0\rangle v$, $v=av_1+bv_2$. The two-by-two matrix 
$\langle 0 | H_{red} | 0\rangle$ is
\beq
\langle 0 | H_{red} | 0\rangle = \begin{pmatrix} -\bar\Omega^*_B \langle 0 | H_B^{-1}|0\rangle  \bar\Omega_B  & \Omega_D \\
\Omega_D & 0 \end{pmatrix} .
\eeq{mm29}
To compute $\langle 0 | H_B^{-1}|0\rangle$ we use $H_B^{-1}=i\int_0^\infty dt \exp{-iH_Bt} $ and 
eqs.~(\ref{mm13}-\ref{mm15}) (without
the term proportional to $ \Im \gamma_L$ because of the phase shift made in going from~\eqref{mm25} to~\eqref{mm26})
\beq
\langle 0 | H_B^{-1}|0\rangle = i \int_0^\infty dt e^{- {\kappa \over 2} |\gamma_L- \sqrt{\bar{n}}|^2  
\left[ t +{2\over \kappa} \left( e^{-\kappa t/2} -1\right)\right]} \equiv {2i\over \beta_B}  .
\eeq{mm30}
In the limit $\bar{n}\gg 1 $ we have $|\gamma_L- \sqrt{\bar{n}}| \approx |\sqrt{\bar{n}}| \gg 1$ so we can evaluate  
integral~\eqref{mm30} using the steepest-descent approximation
\beq
{2\over \beta_B} = \int_0^\infty dt e^{- {\kappa \over 2} |\gamma_L- \sqrt{\bar{n}}|^2  
\left[ t +{2\over \kappa} \left( e^{-\kappa t/2} -1\right)\right]}\approx \int_0^\infty dt 
e^{- |\gamma_L- \sqrt{\bar{n}}|^2 (\kappa t/2)^2 /2} = {\sqrt{2 \pi} \over \kappa |\gamma_L- \sqrt{\bar{n}}|}  .
\eeq{mm31}
Notice that $\beta_B$ is not the intrinsic width of the bright level! Eq.~\eqref{mm10} identifies the decay rate of the survival
probability as $\kappa \bar{n}$. The difference between the two becomes significant for large photon number 
$\bar{n}$, since $\beta_B$ scales with $\sqrt{\bar{n}}$ instead of $\bar{n}$.

The eigenvalues of eq.~\eqref{mm29} are obtained by solving the quadratic equation 
$E(E+2i \bar\Omega^*_B  \bar\Omega_B/\beta_B ) -\Omega_D^2=0$
\beq
E=- i {\bar\Omega^*_B  \bar\Omega_B \over \beta_B} \left[ 1 \pm \sqrt{ 1- {\Omega_D^2 \beta_B^2 \over \bar\Omega_B^2 \bar\Omega_B^{*\, 2} } } \right] .
\eeq{mm32}
For $1\gg \epsilon \gg \eta$, $\epsilon \equiv |\bar\Omega_B|/\beta_B$, $\eta \equiv |\Omega_D/\bar\Omega_B|$ we recover 
the hierarchy (cfr.~\cite{regimes})
\beq
iE_+ = 2 \beta_B \epsilon^2 , \qquad iE_-= {\beta_B \over 2} \eta^2 , \qquad \beta_B \gg i E_+ \gg iE_- .
\eeq{mm33}
In general the norm of the $ |G\rangle; |B\rangle; |D\rangle$ system has 3 decay constants. The condition that the system starts out in a dark period has eliminated the fastest decay eigenvalue leaving the rates~\eqref{mm33} as determining the long time to the next jump. The only decay term in this model is $\beta_B$ ; the cavity induced lifetime of $|B\rangle$. Therefore,
the next quantum jump will be accompanied by an occupation of $ |B\rangle$ and a turn on of the measuring apparatus tuned to 
$|B\rangle$.
This comprises our calculation of the dark period observed in reference 7.

\subsection{What is $\beta_B$?}\label{beta}
As we mentioned earlier, $\beta_B$ cannot be identified with the width of the bright level so it is interesting to try to 
understand what it does represent. A clue comes from applying second-order time-dependent perturbation theory to
a two-level atom $(|B\rangle,  |G\rangle)$. The effective Hamiltonian for this system is
\beq
  H_{eff}  = \begin{pmatrix} H_B & \bar\Omega_B   \\ \bar\Omega_B^* & 
  H_{\not{B}}  \end{pmatrix},
\eeq{mm33a}
with the same expressions for $H_B$,$H_{\not{B}}$ as in~\eqref{mm25}.
 If the system starts in the state $|G\rangle |0\rangle $ then the survival amplitude of this state after
 a time $T$ is
 \beq
 A(T)= 1- \int_0^T dt \int_0^t ds  |\bar{\Omega}|^2 \langle 0 | e^{-iH_B(t-s)} |0\rangle = 1 +
 iT|\bar{\Omega}|^2  \langle 0| H_B^{-1} | 0\rangle
 + |\bar{\Omega}|^2 \langle 0 | {e^{-iHT} -1 \over H_B^2} | 0\rangle .
 \eeq{mm33b} 
 
 Heuristically, one can understand the term proportional to $|\bar{\Omega}|^2$ in this equation as follows: at time $s$ the system undergoes a Rabi transition from $|G\rangle$ to
 $|B\rangle$, with the cavity still in the state $|0\rangle$. When the atom is in $|B\rangle$, $|0\rangle$ is no longer
 an eigenstate of the new cavity Hamiltonian, $H_{B}$, so it starts evolving in time. At time $t$ the cavity is in the state
 $e^{-iH_B(t-s)}|0\rangle$. The overlap with the vacuum is $\langle 0 | e^{-iH_B(t-s)} |0 \rangle$. 
 So the sequence of events
 in which the state $|G\rangle |0 \rangle$ Rabi flops to $|B\rangle |0 \rangle$ at time $s$, evolves until  time $t$, 
 then returns to
 $|G\rangle |0 \rangle$ with a Rabi flop at time $t$ has an amplitude $|\bar{\Omega}|^2 \langle 0 | e^{-iH_B(t-s)} |0\rangle$.
The integration in $t$ and $s$ occurs because in quantum mechanics  it is amplitudes, not probabilities, that  sum 
coherently.
All of this suggests that $2/\beta_B$ is the amplitude for the long term survival of the cavity state $|0\rangle$ 
when the atom is in $|B\rangle$. Notice that the cavity couples to the atom, so $|0\rangle$ is the ground state only
when the atom is in $|G\rangle$. When the atom is in $|B\rangle$, the cavity coupling to photons changes and 
$|0\rangle$ becomes an excited state that is no longer stationary.

\subsection{Limits of validity of the approximation}\label{lim-val}
We computed the lifetime of the dark period using first order perturbation
theory applied to the Hamiltonian~\eqref{mm27}, where we took as perturbation 
the operator $\bar{\Omega}^*_B (E-H_B)^{-1} \bar{\Omega}_B$. First order
perturbation theory is justified as long as corrections coming from the second
order are small. If we perturb an eigenstate $|0\rangle $ with a  perturbation
$V$, then the  condition for the validity of the approximation is
\beq
|\langle 0 | V | 0 \rangle|  \gg 
\sum_n \left| {\langle 0 | V | n \rangle \langle n | V |0 \rangle \over E_n - E_0 }\right| .
\eeq{mm33c}
the RHS can be estimated by replacing the sum with the largest summand, which
typically is the element with the smallest denominator, i.e. the transition to
the energy level closest to $E_0$. In our case the unperturbed energy is $E_0=0$
and the closest level is $a^\dagger |0 \rangle$, whose energy is
$E_1=-i\kappa/2$. So equation~\eqref{mm33c} can be approximated by
\beq
2|\bar{\Omega}_B|^2/\beta_B \gg  \left| {\langle 0 | V | 1 \rangle \langle 1 | V |0 \rangle \over E_1 - E_0 }\right| \sim
2\left[ 2|\bar{\Omega}_B|^2 /\beta_B \right]^2 /\kappa .
\eeq{mm33d}
This gives a rough estimate for the regime of validity of first order perturbation theory.
\beq
|\bar{\Omega}_B|^2/\beta_B \ll \kappa.
\eeq{mm33e}
This equation  shows in particular that the limit $\kappa \rightarrow 0$,
$\kappa \sqrt{\bar{n}}=$constant, $|\bar{\Omega}_B|=$constant is beyond the
limits of our approximation. For large
dispersion the multiscale approximation enables a description with the
weaker restriction: $\Omega^2 < \beta_B^2$ .

\section{Multiscale approximation}\label{multiscale}
We have already obtained one new result in our analysis, namely the explicit relation between 
the lifetimes of various transmon states 
and the parameters $\kappa,\chi,\sqrt{\bar{n}},\Omega_{B,D}$ characterizing the transmon-cavity system. The relation 
is given by eqs.~(\ref{mm31},\ref{mm32}). So far we have also confined ourselves to studying the evolution of the 
system for times $t\geq 1/\kappa$. This is the regime where the SSE applies and heterodyne detection is efficient. 
Yet, the Lindblad equation can describe single photon detection, so it can be
applied to studying the dynamics of our system for times shorter than $1/\kappa$. To study the short-time
regime we need to go beyond perturbation theory and use instead a {\em multiscale approximation}.

Let us write the time evolution of a state $|\psi(t)\rangle$ under the Hamiltonian $H=H_0+ H_I$ as
\beq
|\psi(t)\rangle = e^{-iH_0t}|\psi(0)\rangle -i e^{-iH_0t} \int_0^t ds e^{iH_0s} H_I |\psi(s)\rangle.
\eeq{mmm1}
We will specialize this equation to the two-level atom with Hamiltonians
\beq
H_0=\begin{pmatrix} H_B  & 0   \\ 0 & 
  H_{\not{B}} \end{pmatrix} , \qquad H_I= \begin{pmatrix} 0 & \Omega   \\ \Omega^* & 0 \end{pmatrix} .
\eeq{mmm2}
The state vector, written in a Fock basis for the oscillators $c,c^\dagger$ is
\beq
|\psi(t)\rangle =   \sum_n \begin{pmatrix} C_{B,n} |n\rangle \\ C_{G,n}|n\rangle \end{pmatrix} \equiv 
\begin{pmatrix} |B(t)\rangle \\ |G(t) \rangle \end{pmatrix}.
\eeq{mmm3}
We choose as initial condition $C_{G,0}=1$ with all other coefficients equal to zero so that eq.~\eqref{mmm1} becomes
\beq
  \begin{pmatrix} |B(t)\rangle  \\ |G(t) \rangle \end{pmatrix} = 
 \begin{pmatrix}  -i\Omega \int_0^t ds e^{-i(t-s)H_B} |G(s)\rangle \\ e^{-itH_{\not{B}}} |0\rangle -i \Omega^* 
 \int_0^t ds e^{-i(t-s)H_{\not{B}} }|B(s)\rangle \end{pmatrix} .
 \eeq{mmm4}
The lowest entry in the r.h.s. of this equation simplifies when the detuning $\chi$ is larger than any other scale in the
system; in particular, when it is much larger than $\kappa$.  When $\chi \gg \kappa$ we can use
\beq
\lim_{\chi \rightarrow \infty} \overline{\langle n | e^{-i(t-s)H_{\not{B}}} |B(t)\rangle} \approx 
\overline{e^{i(t-s)\chi n} \langle n | B(t) \rangle} =0
\mbox{ for } n > 0,
\eeq{mmm5}
to neglect all terms in $|G(t)\rangle$ except the $n=0$ one. Here the overbar denotes averaging over time intervals 
$\Delta t  \gg 1/\chi n$. The ansatz $|G(t)\rangle = C_{G,0}(t)|0\rangle$ allows for
a closed, self-consistent solution of eq.~\eqref{mmm4}
\beq
\begin{pmatrix} |B(t)\rangle  \\ |G(t) \rangle \end{pmatrix} = 
 \begin{pmatrix}  -i\Omega \int_0^t ds e^{-i(t-s)H_B} C_{G,0}(s)|0\rangle \\  |0\rangle -\Omega\Omega^* |0\rangle \langle 0 |
 \int_0^t ds e^{-i(t-s)H_{\not{B}} }  \int_0^s dw e^{-i(s-w)H_B} C_{G,0}(w)|0\rangle\end{pmatrix} .
 \eeq{mmm6}
By taking the derivative w.r.t. $t$ of the lowest component in the r.h.s. of this equation and using eqs.~\eqref{mm13},
\eqref{mm14},\eqref{mm15} with $\gamma_L=0$,we get 
\beq
{d\over dt} C_{G,0}(t)= -\Omega\Omega^* \int_0^t dw \langle 0 | e^{\alpha(t-w) a^\dagger + \beta(t-w) }
|0\rangle  C_{G,0}(w)= -\Omega\Omega^* \int_0^t dw e^{ \beta(t-w) }  C_{G,0}(w).
\eeq{mmm7}
For $t\gg 1/\kappa \sqrt{\bar{n}}$ we get the same lifetime as in first-order time-independent perturbation theory. 
Specifically, when the shortest time scale is $\kappa \sqrt{\bar{n}}$, we can use the Gaussian approximation~\eqref{mm31} 
to obtain $dC_{G,0}(t)/dt= -\gamma C_{G,0}(t)$, $\gamma= 2\Omega\Omega^*/\beta$.

The norm $\langle \psi (t) | \psi(t) \rangle = \langle G(t) | G(t) \rangle + \langle B(t) | B(t) \rangle $ can be computed 
explicitly, for $t\gtrsim 1/\kappa \sqrt{\bar{n}}$, in the regime $\bar{n} \gg 1$, where 
$\kappa \sqrt{\bar{n}} \gg \kappa \bar{n}^{1/3}$. 
For  $t\gtrsim 1/\kappa \sqrt{\bar{n}}$ we immediately find  $\langle G(t) | G(t) \rangle = \exp(-2\gamma t)$, so the only 
nontrivial term to compute is $\langle B(t) | B(t) \rangle $.
By using again eqs.~\eqref{mm13}, \eqref{mm14}, \eqref{mm15} and with the change of variable of integration $w=t-s$, we find 
\beq
|B(t)\rangle = -i\Omega \int_0^t dw e^{\alpha(w)a^\dagger + \beta(w) }C_{G,0}(t-w) |0 \rangle. 
\eeq{mmm8}
Using the 
approximation $C_{G,0}(t)=\exp(-\gamma t)$, we find also 
\bea
\langle B(t) | B(t) \rangle &=&  e^{-2\gamma t} \Omega \Omega^* \int_0^t dw \int_0^t dw' e^{\gamma (w +w')}
\langle 0 | e^{\beta^*(w') + \alpha^*(w')a} e^{\beta(w) + \alpha(w)a^\dagger} |0 \rangle  \nonumber \\
&=& e^{-2\gamma t} \Omega \Omega^* \int_0^t dw \int_0^t dw' e^{\gamma (w +w')+
 \beta^*(w') + \beta(w) + \alpha(w)\alpha^*(w')} .
\eea{mmm9}
When $t\gtrsim \kappa^{-1} \bar{n}^{-1/3}$, this norm decays as $\exp(-2\gamma t)$, so we recover the first-order result
for the survival probability of the state.

The multiscale approximation gives an interesting new behavior for the survival probability of the state at intermediate 
times: $\kappa^{-1} \bar{n}^{-1/3} \gtrsim t \gtrsim \kappa^{-1} \bar{n}^{-1/2}$.
In this regime, by expanding to quadratic order in $w,w'$ we find $\beta^*(w') + \beta(w) + \alpha(w)\alpha^*(w')=
-(\kappa^2 \bar{n} /8)(w^2 + w'^2 -2ww')$. The change of variables $w=x+y/2$, $w'=x-y/2$  transforms the integral in
eq.~\eqref{mmm9} into
\beq
\int_0^t dx \int_{-t}^t dy e^{2\gamma x -\kappa^2 \bar{n} y^2/8 +2 \beta(x) +\alpha(x)\alpha^*(x) + O(y^3)} 
\approx {2\over \kappa} \sqrt{ 2\pi \over \bar{n}} \int_0^t dx e^{2\gamma x + 2 \beta(x) +\alpha(x)\alpha^*(x)} .
\eeq{mmm11}
The  expansion of $2 \beta(x) +\alpha(x)\alpha^*(x)$ begins at order $x^3$ and gives
\beq
\langle B(t) | B(t) \rangle \approx e^{-2\gamma t} \Omega \Omega^* {2\over \kappa} \sqrt{ 2\pi \over \bar{n}} 
\int_0^t dx e^{2\gamma x - \kappa^3 \bar{n} x^3/12}.
\eeq{mmm12}

Now we are ready to put all together: when $\kappa^{-1} \bar{n}^{-1/3} \gtrsim t \gtrsim \kappa^{-1} \bar{n}^{-1/2}$ 
 the time derivative of the norm of the state is approximated by 
\beq
{d\over dt} \langle \psi(t)| \psi(t) \rangle = -2\gamma \langle \psi(t)| \psi(t) \rangle + 
\Omega \Omega^* {2\over \kappa} \sqrt{ 2\pi \over \bar{n}} 
 e^{- \kappa^3 \bar{n} t^3/12}.
\eeq{mmm13}

We can show that ${d\over dt} \langle \psi(t)| \psi(t) \rangle \leq 0$ as follows. Eq.~\eqref{mmm13} simplifies to
\beq
{d\over dt} \langle \psi(t)| \psi(t) \rangle= -2\gamma e^{-2\gamma t}\left[ 1+ 2\gamma 
\int_0^t dx e^{2\gamma x - \kappa^3 \bar{n} x^3/12}\right] +2\gamma e^{- \kappa^3 \bar{n} t^3/12} .
\eeq{mmm14}
This expression vanishes at $t=0$ while at $t>0$ is negative, because by multiplying the l.h.s. of~\eqref{mmm14}
by $\exp(2\gamma t)$ and taking its derivative we get
\beq
{d \over dt} \exp(2\gamma t) {d\over dt} \langle \psi(t)| \psi(t) \rangle  = -4\gamma^2 e^{2\gamma t- \kappa^3 \bar{n} t^3/12}+2\gamma \left(2\gamma -{\kappa \bar{n} t^2 \over 4 } \right)e^{2\gamma t- \kappa^3 \bar{n}t^3/12 } \leq 0.
\eeq{mmm15}

\section{Evolution under continuous observation of the unshifted photon field}\label{unshifted}
In this case the evolution equation is
\beq
  \begin{pmatrix} |B(t)\rangle  \\ |G(t) \rangle \end{pmatrix} = 
 \begin{pmatrix}  e^{-itH_B}|B(0)\rangle -i\Omega \int_0^t ds e^{-i(t-s)H_B} |G(s)\rangle \\ e^{-itH_{\not{B}}} |G(0)\rangle -i \Omega^* 
 \int_0^t ds e^{-i(t-s)H_{\not{B}} }|B(s)\rangle \end{pmatrix} ,
 \eeq{mmm16}
 with Hamiltonians
 \beq
H_B=-i{\kappa \over 2} c^\dagger c + {\kappa \over 2i} \sqrt{\bar{n}} (c -c^\dagger), \qquad
H_{\not{B}}= -\chi c^\dagger c + H_B.
\eeq{mmm17}
It is convenient to introduce shifted oscillators $a=c-A$, $b^\dagger=c^\dagger-B^\dagger$. They obey canonical 
commutation relations: $[a,b^\dagger]=1$, but they are \underline{not} Hermitian conjugate to each other. By choosing
\beq
A= -{\kappa \sqrt{\bar{n}}/2i \over \chi +i\kappa/2}, \qquad B= {\kappa \sqrt{\bar{n}}/2i \over \chi +i\kappa/2},
\eeq{mmm18}
we can diagonalize the Hamiltonian 
\beq
H_{\not{B}}=-(\chi +i\kappa/2)[b^\dagger a - n_0], \qquad n_0={\kappa^2 \bar{n}/4 \over (\chi + i\kappa/2)^2}. 
\eeq{mmm19}
A basis of linearly independent but 
\underline{not} orthogonal eigenvectors of $H_{\not{B}}$ is $|n\rangle= C_n  (b^{\dagger})^n | 0 \rangle $. We won't need to
specify the constants $C_n$. 
Next we write the vector $|B(t)\rangle$ as $|B(t)\rangle=\sum_n C_{B,n}(t)|n\rangle $ and write 
the term $\int_0^t ds e^{-i(t-s)H_{\not{B}} }|B(s)\rangle$ in equation~\eqref{mmm16} as
\beq
 \int_0^t ds e^{-i(t-s)H_{\not{B}} }|B(s)\rangle =\sum_n\int_0^t ds e^{i(\chi +i\kappa/2)(n-n_0)(t-s)}C_{B,n}(s) |n\rangle .
 \eeq{mmm20}
 In the limit $\chi \rightarrow \infty$ all terms with $n\neq 0$ in the sum in~\eqref{mmm20} vanish. When we choose the 
 initial condition $|G(0)\rangle=|0\rangle$, $|B(0)\rangle=0$, we can use as we did earlier the ansatz 
 $|G(t)\rangle= C_{G,0}(t)|0\rangle$. 
 In the limit $\chi\rightarrow \infty$ the Hamiltonian $H_B$ becomes 
 \beq
 H_B= -i{\kappa \over 2} b^\dagger a + {\kappa \over 2i} \sqrt{\bar{n}} (a -b^\dagger) + O(1/\chi).
 \eeq{mmm21}
 By using the same manipulations that we used in section 4, we arrive at the equation
 \bea
\left[{d\over dt} +i(\chi +i\kappa/2)n_0\right]C_{G,0}(t) &=& -\Omega\Omega^* \int_0^t dw \langle 0 | e^{\alpha(t-w) b^\dagger + 
\beta(t-w) } |0\rangle  C_{G,0}(w) \nonumber \\ &=& -\Omega\Omega^* \int_0^t dw e^{ \beta(t-w) }  C_{G,0}(w),
\eea{mmm22}
 with
 \beq
 \beta(t)= {\kappa \bar{n} \over 2} \left( 1-i {\kappa/2\over \chi +i \kappa/2}\right)^2 \left[ t + {e^{-\kappa t/2}-1 \over \kappa/2}
 \right]  -i {\kappa^2 \bar{n} \over 4} {2\chi +i\kappa/2 \over (\chi + i\kappa/2)^2} t = {\kappa^2 \bar{n} \over 8} t^2 + 
 O(1/\chi) .
 \eeq{mmm23}
 So, following the computation in eq~\eqref{mmm7} we get, up to terms $O(\Omega\Omega^*/\chi)$, 
 \beq
 C_{G,0}(t)= e^{\Gamma t}, \qquad \Gamma=-i(\chi +i\kappa/2)n_0 - {\sqrt{2\pi\over \kappa^2 \bar{n}}} \Omega\Omega^* =
 -i{\kappa^2 \bar{n}/4 \over (\chi + i\kappa/2)} - {\sqrt{2\pi\over \kappa^2 \bar{n}}} \Omega\Omega^*,
  \eeq{mmm24}
 consistent with eq.~\eqref{lts}
 
 \section*{Conclusions}\label{conc}
 When atomic fluorescence is conditioned on the observation of the next quantum jump there appear large experimentally measurable correlations in the time between events. We have carried this formalism over to solid state systems such as transmon qubits that are dispersively coupled to a readout cavity. When the cavity is treated quantum mechanically the next photon formalism again yields the possibility for long dark periods in a driven system; as well as new correlations for short time readout. Coupling of a transmon to a driven cavity also leads to a finite lifetime of the discrete transmon levels. Single photon detection has not been applied to solid state qubits because this technique is difficult to implement at microwave frequencies. Instead signal acquisition via a heterodyne technique is currently used. The formulation of a quantum measurement conditioned on heterodyne readout involves assumptions on noise and time scales which lead to a stochastic Schr\"odinger equation which we argue is not fundamental as compared to the next jump formalism that we develop. The practical realization of that approach hinges upon advanced instruments that are capable of measuring single microwave photons.  Progress in this direction could make the extension of our theory of fluorescence to solid state systems feasible. In this case new measurement correlations such as rapid readout will become available.  
 Underlying the new class of quantum correlations which we study is an effective
non-Hermitian Hamiltonian which applies to non-stationary/ non-Poissonian
quantum systems. Extending the method of reference~\cite{pp} we showed that a density
matrix evolving under both nonstationary drive and null measurement factors into
a product of wave functions each evolving coherently according to the effective
Hamiltonian. 
 \subsection*{Acknowledgments} 
 We wish to acknowledge valuable discussions with Hong Wen Jiang. 

\setcounter{section}{0}
\renewcommand{\thesection}{\Alph{section}}
\section{On heterodyne detection and the SSE, again}\label{het2}
\setcounter{equation}{0}
\renewcommand{\theequation}{A.\arabic{equation}} 
There is an apparent disagreement in the literature regarding which SSE describes heterodyne photodetection. The last 
item listed in section~\ref{meas} shows that the disagreement is only apparent. 
Notably, eq.~(4.18) of ref.~\cite{wiseman-milburn} is (using the notation of references~\cite{wiseman,wiseman-milburn} where $\kappa=1$ and $\langle\langle \dot\xi(t) \dot\xi(t')\rangle\rangle =\delta(t-t')$)

\bea
i\dot\chi &=& \left( i[\dot\xi e^{-i\phi} + I]c + H_A+H_R -{i\over 2}c^\dagger c \right)\chi, \nonumber \\
I &=& {(  \chi , (c +c^\dagger ) \chi ) \over ( \chi , \chi ) } ,
\eea{3-15-1}
so it differs from our eq.~\eqref{mm18}, hence from eq.~(41) of 
\cite{wiseman}, because it contains a classical drift term proportional $I$. 
The redefinition $\chi= \exp(Bc+i D)\psi$  transforms eq.~\eqref{3-15-1} 
into~\eqref{mm18} when $\dot B = I +B/2$, $\dot D= AB$. For $t\gg 1$ $\psi$ becomes proportional to the coherent state
$| -2iA )$ so $\chi$ becomes equal to $\exp(-2iAB +iD) \psi$. So the conditioned states obtained from eq.~\eqref{mm18} 
and the SSE~\eqref{3-15-1} are actually the same. Moreover, the expectation values of the quadrature 
$c+c^\dagger$  also agree, because
\beq
I={(  \chi , (c +c^\dagger ) \chi )  \over (  \chi , \chi ) }={(  e^{-2iAB+iD}\psi , (c +c^\dagger ) e^{-2iAB+iD}\psi )  \over 
(  e^{-2iAB+iD}\psi , e^{-2iAB+iD}\psi ) }={(  \psi , (c +c^\dagger ) \psi )  \over 
(  \psi , \psi )} .
\eeq{3-15-2}
\section{Another digression into continuous measurement theory}\label{appB}
\setcounter{equation}{0}
\renewcommand{\theequation}{B.\arabic{equation}} 
This appendix presents a simple model of photodetection, that applies to the 3-level atom. It is meant to
show why we did not need to worry about the detector in~\cite{pp} and also to show how to apply the general formalism
outlined in section~\ref{meas} to a concrete model of detector, whose resolution time for a measurement will be denoted
by $1/ \gamma$.

Consider an atom in interaction with the EM field monitored by a $4\pi$ photodetector. The Hamiltonian of the atom+radiation
system is $H$, while the Hamiltonian of the photodetector is (in the Schr\"odinger representation) 
\beq
H_D=\sum_{aI} H_{D\,Ia}, \qquad H_{D\,Ia} = \sum_{Ia}\int d^3{\bf x} J_{Ia}({\bf x},t)\left[  |1 \rangle_{Ia} \langle 0 |_{Ia} {\bf A}^+({\bf x})\cdot {\bf D} + h.c. \right]
\eeq{mm34}
The space- and time-dependent coupling constants $J_{Ia}$ are characteristic functions of the region $V_I \times T_a$,
$V_I=\{{\bf x}: x^i_I-\Delta/2 < x^i < x^i_I + \Delta/2\}$ , $T_a=\{t: t_a - T/2 < t < t_a + T/2\} $ ($J_{Ia}=1$ inside and $J_{Ia}=0$ outside) or a smooth version
thereof. The constant vector $\bf D$ characterizes the strength of the interaction and ${\bf A}^+$ is the positive-frequency part 
of the EM vector potential (in Coulomb gauge). Each of the Hamiltonians $H_{D\, Ia}$ describes a pixel  of the photodetector,
centered in ${\bf x}_I$ and of size $\Delta^3$. The pixel records the presence of a photon in the volume  $V_I$
during the time interval $T_a$ by changing the state of some recording system from its initial value $|0\rangle_{Ia}$ to 
$|1\rangle_{Ia}$. If $V_I \times T_a$ is sufficiently small the probability of erasure or multiple recording can be made arbitrarily 
small. What the detector does is that it takes pictures of the regions $V_I$ at times $t_a$ and files them away 
(i.e. it establishes a permanent record).

We can compute the evolution of the system {\em atom + EM field + photodetector} using first-order time-dependent perturbation theory, starting from a state with no photons and with the ``recorder'' (ancilla) Hilbert space set to $\prod_{Ia} |0\rangle_{Ia}$ as
\beq
|\psi(t)\rangle = \prod_{Ia} \left(1- i\int dt e^{iHt} H_{D\,Ia} e^{-iHt} \right) |\psi(0)\rangle.
\eeq{mm35}
If there is no detection up to a time $t_n-T/2$, the ancilla is still in the state $\prod_{Ia} |0\rangle_{Ia}$. So the state of the 
system is $|\psi(t)\rangle =|\phi(0) \rangle \prod_{Ia} |0\rangle_{Ia}$, where $|\phi(0)\rangle $ is the state vector of atom + EM field. 
(In interaction representation we factor out the free Hamiltonian evolution; here ``free'' means the complete Hamiltonian 
without the detector). The state vector at time $t_n + T/2$ is then
\beq
|\psi(t_n+T/2)\rangle = \left[ \prod_K |0\rangle_{Kn} -i \sum_I \prod_{K\neq I} |0\rangle_{nK} |1\rangle_{nI} 
\int_{t_n-T/s}^{t_n+T/s}dt \int d^3{\bf x} J_{Ia}({\bf x},t) {\bf A}^+({\bf x},t)\cdot {\bf D} \right] | \phi(0)\rangle.
\eeq{mm36}
The EM fields are now in the interaction representation (they evolve with $\exp(iHt) {\bf A} \exp(-iHt)$). 
We take now $\Delta$ and $T$ infinitesimal so that eq~\eqref{mm36} can be approximated as
\beq
|\psi(t_n+T/2)\rangle = \left[ \prod_K |0\rangle_{Kn} -i \Delta^3 T\sum_I \prod_{K\neq I} |0\rangle_{nK} |1\rangle_{nI} 
  {\bf A}^+({\bf x_I},t_n)\cdot {\bf D} \right] | \phi(0)\rangle.
\eeq{mm37}
To go back to 
the Schr\"odinger picture we must multiply the vector $|\psi(t_n+T/2)\rangle$ by $\exp[-iH(t_n+T/2)]$. 
The states $|0\rangle_{Ia}, |1\rangle_{Ia}$ are orthonormal so the probability of recording a detection at time $t_n+T/2$ is
\beq
P(t_n+T/2) = T^2 \Delta^6 \sum_I \left[ \tr {\bf A}^-({\bf x_I},t)\cdot {\bf D}^* {\bf A}^+({\bf x_I},t)\cdot {\bf D} \rho(0)\right], \qquad
\rho(0)\equiv |\phi(0)\rangle \langle \phi(0)| .
\eeq{mm38}
We rescale the dipole interaction as ${\bf D}=\Delta^{-3/2} T^{-1/2} {\bf d}$, approximate the sum over $I$ with an integral and return to the Schr\"odinger representation to write 
\beq
P(t_n+T/2)= T  \int d^3  {\bf x} \left[ \tr  {\bf A}^-({\bf x})\cdot {\bf d}^* {\bf A}^+({\bf x})\cdot {\bf d} \rho(t_n-T/2) \right],
\eeq{mm39}
This formula can be easily generalized to the case of several dipoles that can be chosen e.g. to average to a Kronecker delta
$d^{*\, i} d^j \rightarrow \sum_\alpha d_\alpha^{*\, i} d^j_\alpha = \delta^{ij} \gamma$. By taking $T$ infinitesimal we can then write formula~\eqref{mm39} as
\beq
P(t+dt )= \gamma dt  \int d^3  {\bf x} \left[ \tr  {\bf A}^-({\bf x})\cdot  {\bf A}^+({\bf x}) \rho(t) \right],
\eeq{mm39a}
This is exactly the ``measurement'' term that appears in the Lindblad operator. The probability of not recording anything 
in the time interval $dt$ is ($P+Q=1$)
\beq
Q(t+dt) = Q(t) -\gamma dt  \int d^3  {\bf x} \left[ \tr  {\bf A}^-({\bf x})\cdot  {\bf A}^+({\bf x}) \rho(t) \right], \qquad Q(t)=\tr \rho(t).
\eeq{mm40}
This equation says that the evolution when no photon is detected by the measurement instrument is governed by the effective 
Hamiltonian
\beq
H_{eff} =H -i {\gamma \over 2}  \int d^3  {\bf x}  {\bf A}^-({\bf x})\cdot  {\bf A}^+({\bf x}).
\eeq{mm41}
States with photons are short lived: if their average photon number is $N$ their lifetime is $1/N\gamma $. That of course means 
simply that our photodetector cannot resolve times shorter than $1/\gamma$ but it is good at detecting photons over a 
time longer than $1/\gamma$. States with no photons can be long-lived. Their
lifetime can be computed using time-independent perturbation theory to second order. Calling $\mathbf{P}$ the projection over the 
zero-photon state and $\mathbf{Q}$ its orthogonal complement and writing $H_{eff}=H_0+H_I$, with $H_I$ the atom-EM interaction term,  we must diagonalize the   matrix
\beq
H_{eff\; 0} (E) =\mathbf{P}H_{eff}\mathbf{P}-\mathbf{P} H_I \mathbf{Q} (\mathbf{Q}H_{eff}\mathbf{Q} -E)^{-1} \mathbf{Q}H_I \mathbf{P}.
\eeq{mm42}
For the 3-level atom this was done e.g, in~\cite{pp}.
The largest decay width is associated with the transition $|B\rangle \rightarrow | G\rangle$. To  lowest order in perturbation theory 
it is given in terms of the atomic dipole matrix elements ${\bf \mu}$ as~\footnote{The form factor appearing in
this equation has appeared previously in other papers on continuous measurement, e.g.~\cite{fnp}.}
\beq
\Gamma_B  \equiv=\Im E = -\Im \int {{d\!\!{^-}}^3{\bf k} \over 2|{\bf k}|}{ |{\bf \mu}|^2 \over |{\bf k}| -i\gamma/2 -\omega_{BG} }=
-\int {{d\!\!{^-}}^3{\bf k} \over 2|{\bf k}|} { |{\bf \mu}|^2 \gamma/2 \over (|{\bf k}|  -\omega_{BG})^2 + \gamma^2/4}.
\eeq{mm43}
The usual formula for the decay rate, which was used in~\cite{pp}, is the limit $\gamma\rightarrow 0$ of eq.~\eqref{mm43}.
The change of variables $|{\bf k}| =\gamma x/2 + \omega_{BG}$ shows that integral~\eqref{mm43} differs from the ``unobserved''
value $\Gamma_B^0\equiv\lim_{\gamma\rightarrow 0} \Gamma_B$ by terms $O(\Gamma_B^0\gamma/\omega_{BG})$.
In conclusion, the presence of the detector does not change significantly the decay rate, as long as the rate of detection
$\gamma$ is much smaller than the difference in frequencies between atomic levels.

\bibliographystyle{unsrt}
 
\end{document}